%% file: main.tex
\documentclass[sigconf,nonacm]{acmart}
\AtBeginDocument{%
  }

\usepackage{amsmath,amsthm}
\usepackage{mathtools}
\usepackage{bm}

\usepackage{graphicx}
\graphicspath{{figures/}}

\usepackage{enumitem}
\usepackage{xcolor}
\usepackage{xspace}
\newtheorem{theorem}{Theorem}
\newtheorem{lemma}[theorem]{Lemma}
\newtheorem{proposition}[theorem]{Proposition}
\newtheorem{corollary}[theorem]{Corollary}
\newtheorem{remark}[theorem]{Remark}
 
\providecommand{\sign}{}
\let\sign\relax
\DeclareMathOperator{\sign}{sign}
\providecommand{\Lq}{L_{q}}
\providecommand{\La}{L_{\alpha}}
\providecommand{\Jbar}{\bar{J}}
\providecommand{\E}{\mathbb{E}}
\providecommand{\indicator}{\mathbf{1}}
\providecommand{\R}{\mathbb{R}}
\renewcommand{\Pr}{\mathbb{P}}

\begin{document}

\title[One AI Signal, Many Human Judgments]{One AI Signal, Many Human Judgments: A Bayesian Cascade Analysis of AI-based Credibility Indicators in Online Information Spread}

\author{Zhuoran Lu}
\affiliation{%
  \institution{Purdue University}
  \city{West Lafayette}
  \state{Indiana}
  \country{USA}}
\email{zhuoranlu33@gmail.com}

\author{Weilong Wang}
\affiliation{%
  \institution{St John's University}
  \city{New York}
  \state{New York}
  \country{USA}}
\email{weilong.wang@stjohns.edu}

\author{Yangyang Yu}
\affiliation{%
  \institution{Stevens Institute of Technology}
  \city{Hoboken}
  \state{New Jersey}
  \country{USA}}
\email{yyu44@stevens.edu}

\author{Xinru Wang}
\affiliation{%
  \institution{Singapore-MIT Alliance for Research and Technology}
  \city{Singapore}
  \country{Singapore}}
\email{xinru.wang@smart.mit.edu}

\author{Zhuoyan Li}
\affiliation{%
  \institution{Purdue University}
  \city{West Lafayette}
  \state{Indiana}
  \country{USA}}
\email{li4178@purdue.edu}

\author{Zhiwei Liu}
\affiliation{%
  \institution{University of Manchester}
  \city{Manchester}
  \country{United Kingdom}}
\email{zhiwei.liu@manchester.ac.uk}

\author{Sophia Ananiadou}
\affiliation{%
  \institution{University of Manchester}
  \city{Manchester}
  \country{United Kingdom}}
\email{sophia.ananiadou@manchester.ac.uk}

\renewcommand{\shortauthors}{Lu et al.}

\begin{abstract}
Social media platforms increasingly use AI-based credibility indicators to help
users judge misinformation. Unlike individual human-AI decision-making, these
indicators are embedded in information spread: users see both an AI prediction
and earlier judgments shaped by the same AI, and their own judgments may then
enter the public history. Yet how to analytically characterize this process remains under-explored. We therefore introduce a social-learning lens for this setting by
extending the classical Bayesian cascade model with the AI indicator as a shared
public signal. The resulting Gateway condition compares the evidence from the AI
prediction with users' private impressions. Through this view, we show that AI changes what public history means. Crowd
agreement may reflect accumulated independent human evidence, or repeated
dependence on the same AI prediction. This creates a preservation--correction
trade-off: stronger reliance on AI can preserve correct predictions, but can
also lock in incorrect ones by blocking corrective private impressions. We
calibrate the model using human-subject data on news veracity judgments. Although
the AI outperforms human users, the average user weights it below her own
impression but above several peer judgments, while individual users vary from discounting
the AI to relying on it enough to cascade. Simulations
show that over-reliance on a weak AI is especially harmful, and that diversifying
AI signals across users can better keep the crowd informative. We conclude with
implications for understanding human-AI interaction in information spread and
designing misinformation interventions.

\end{abstract}



\maketitle

\input{sections/01-introduction}
\input{sections/02-related-work}
\input{sections/03-model}
\input{sections/04-dynamics}
\input{sections/05-main-results}
\input{sections/06-empirical-calibration}
\input{sections/07-simulation}
\input{sections/08-discussion}

\input{sections/09-conclusion}
\begin{acks}
We are grateful to the anonymous reviewers who provided many helpful comments.
Any opinions, findings, conclusions, or recommendations expressed here are
those of the authors alone.
\end{acks}

\bibliographystyle{ACM-Reference-Format}
\bibliography{references}

\appendix
\counterwithin{theorem}{section}
\counterwithin{equation}{section}
\counterwithin{figure}{section}
\counterwithin{table}{section}

\section{Measure-zero boundary events: tie-breaking and the classical edge \texorpdfstring{$\alpha=\tfrac12$}{alpha=1/2}}\label{app:tie-breaking}

This appendix collects the measure-zero events excluded from the main analysis: the tie-breaking convention at the cascade boundary, the alternative random tie-break, and the classical edge $\alpha=\tfrac12$. The upshot, established below, is that the convention bites only at the two measure-zero edges $\alpha\in\{\tfrac12,q\}$ and leaves every main conclusion (cascade genericity, direction selection, the information ceiling, and the spurious-confidence slope) unchanged.

\subsection{The conventional (public) tie-break}

The decision rule of Section~\ref{sec:model} reads $J_i = \indicator\{\ell_i > 0\}$ on the open set $\{\ell_i \neq 0\}$, and is silent on the boundary $\ell_i = 0$. The boundary is a measure-zero event in the interior of the learning region and can be assigned arbitrarily there. It is not measure-zero at the cascade absorbing thresholds, however: when $P_i = +\Lq$, the private impression $s_i = 0$ produces $\ell_i = P_i - \Lq = 0$ exactly, and a tie-breaking convention is needed to make the absorbing-set arguments in Section~\ref{sec:dynamics} clean.

We adopt the convention that ties are broken in the direction of $P_i$, i.e., the user defers to the public log-odds when her private impression exactly cancels them. Concretely, the full decision rule is
\[
J_i \;=\; \begin{cases}
1, & \Pr(\theta=1 \mid s_i, c, H_i) > \tfrac{1}{2}, \\
0, & \Pr(\theta=1 \mid s_i, c, H_i) < \tfrac{1}{2}, \\
\indicator\{P_i \geq 0\}, & \Pr(\theta=1 \mid s_i, c, H_i) = \tfrac{1}{2},
\end{cases}
\]
with $J_i = 1$ by default in the measure-zero event $P_i = 0$. The equivalent log-odds form, with $\ell_i = P_i + \lambda(s_i;q)$, is
\[
J_i \;=\; \indicator\{\ell_i > 0\} \;+\; \indicator\{\ell_i = 0\}\cdot\indicator\{P_i \geq 0\}.
\]

This convention has no effect on the interior of the learning region $|P_i| < \Lq$, where $\ell_i \neq 0$ for both signal realizations $s_i \in \{0,1\}$. It is load-bearing at the boundaries: $P_i = +\Lq$ commits to $J_i = 1$ (the tied private impression $s_i = 0$ is overridden by the positive public log-odds), and $P_i = -\Lq$ commits to $J_i = 0$ (symmetric case). This is exactly what Lemma~\ref{lem:cascade-structure}(b)--(c) requires at equality. Intuitively, when private and public evidence cancel, the user defers to the public component---the same direction a marginal additional public observation would have nudged her.

\subsection{Ties occur only at the two measure-zero edges \texorpdfstring{$\alpha\in\{\tfrac12,q\}$}{alpha in 1/2,q}}\label{app:tie-when}

A tie ($\ell_i=0$ with positive probability) requires the public log-odds to land \emph{exactly} on a boundary, $P_i=\pm\Lq$. With symmetric prior, $P_i = \lambda(c;\alpha) + k\Lq = \pm(\La + k\Lq)$ for some integer $k$ (the net of the revealed pre-cascade impressions). Thus $P_i=\pm\Lq$ with positive probability requires $\La = (1-|k|)\Lq$ for some integer $k$, which for $\alpha\in(\tfrac12,1)$ has exactly two solutions: $\La=\Lq$, i.e.\ $\alpha=q$ ($k=0$), and $\La=0$, i.e.\ $\alpha=\tfrac12$ ($k=1$, the first boundary visit $P_2=\pm\Lq$). For every other $\alpha$---in particular the whole below-Gateway interval $\alpha\in(\tfrac12,q)$ and the strong-AI region $\alpha>q$---cascade entry is \emph{strict} ($|P|>\Lq$ at the moment of absorption, because $\La\neq\Lq$), no tie ever occurs, and \textbf{the tie-breaking convention is irrelevant}. The convention can therefore matter only at the two measure-zero edges, which we treat in turn.

\subsection{The random tie-break}\label{app:tie-random}

The alternative convention resolves each tie by an independent fair coin, $J_i=\indicator\{u_i>\tfrac12\}$ with $u_i\sim\mathrm{Unif}[0,1]$. By the previous subsection it differs from the public rule only at $\alpha\in\{\tfrac12,q\}$.

\emph{At $\alpha=q$.} The first user has $P_1=\pm\Lq$, and a disagreeing private impression ($s_1\neq c$) produces a tie. The public rule sets $J_1=c$, giving the weak-inequality Gateway statement ``$\alpha\geq q\Rightarrow\tau=1$'' (Theorem~\ref{thm:cascade-genericity}). The random rule instead lets user~1 reveal $s_1$ with probability $\tfrac12$, so first-user cascade genericity holds for $\alpha>q$ and at $\alpha=q$ only with probability $\tfrac12$. This shifts a single measure-zero point from the cascade side to the learning side; every statement for $\alpha>q$ is untouched.

\emph{At $\alpha=\tfrac12$.} See the next subsection. In both cases the absorbing \emph{direction} $D_\infty$ keeps the same conditional law---a strict cascade ($|P|>\Lq$) is reached in finite expected time regardless of how the boundary ties are resolved, and the coin is symmetric---so direction selection (Theorem~\ref{thm:direction}), the information-aggregation bound and the spurious-confidence slope $\Lq$ (Theorem~\ref{thm:spurious}) are identical under the two conventions. Only the cascade-formation \emph{time} changes, and only at $\alpha=\tfrac12$.

\subsection{The classical edge \texorpdfstring{$\alpha=\tfrac12$}{alpha=1/2}}\label{app:classical-edge}

When $\alpha=\tfrac12$ the AI prediction contributes zero log-odds, so $P_1=0$: the first user is in the interior and acts on her private impression, $J_1=s_1$, sending $P_2=\pm\Lq$ to a boundary. From there the conventions diverge.

\emph{Cascade-formation time.} Under the \textbf{public} tie-break, the boundary $P_2=\pm\Lq$ is absorbing: every later user with a disagreeing impression is overridden toward the public side, so $\tau=2$, and $J_j=J_1$ for all $j\geq2$---only the first user's impression is ever revealed. Under the \textbf{random} tie-break, a disagreeing impression at the boundary registers with probability $\tfrac12$ (knocking $P$ back to $0$ before it re-absorbs), so $\tau$ takes values in $\{2,3,\ldots\}$ and a geometric number ($\E[\tau-1]<\infty$) of private impressions enter the record. This is the only place the cascade-time law of Proposition~\ref{prop:tau} depends on the convention.

\emph{Why the main conclusions survive.} Either way the revealed prefix is finite in expectation, so the information $I(\Jbar_N;\theta)$ still saturates at an $N$-independent ceiling, and the naive external observer's excess still grows at slope $\Lq$ (Theorem~\ref{thm:spurious}). Under the public rule the consensus copies $J_1$, so the Bayes-correct observer extracts exactly one private impression and the spurious-confidence constant is $O(1)=-\Lq$; under the random rule the constant absorbs the few extra revealed impressions but remains uniformly bounded in $N$. The classical edge thus reproduces, rather than alters, the qualitative content of the main theorems.

\section{Continuous-Valued AI Indicators}\label{app:continuous}

We extend the analysis to continuous-valued AI indicators (e.g., AI-generated credibility indicators on a real-valued scale rather than a binary flag). Let the AI signal $c \in \R$ be the indicator's real-valued output, whose conditional density $f_\theta(c)$ satisfies the strict monotone likelihood ratio property: $\frac{f_1(c)}{f_0(c)}$ is strictly increasing in $c$. The Bayesian update with this $c$ replaces the binary log-likelihood term $\lambda(c; \alpha)$ with the log-likelihood ratio
\[
\Lambda(c) \equiv \log\!\frac{f_1(c)}{f_0(c)}.
\]
The threshold structure of Theorem~\ref{thm:equilibrium} generalizes immediately: the public log-odds at user $i$ become
\[
P_i = \ell_0 + \Lambda(c) + \sum_{j < i} \psi_j(J_j),
\]
and the cascade thresholds remain at $\pm\Lq$. The continuous analog of cascade genericity is:

\begin{theorem}[Continuous-Indicator Immediate Absorption]\label{thm:cont-genericity}
Let $G_q \equiv \Pr(|\Lambda(c)| \geq \Lq)$. Then with probability $G_q$, a cascade forms at the first user. Conditional on $c$ with $|\Lambda(c)| \geq \Lq$, $J_i = \indicator\{\Lambda(c) > 0\}$ for all $i \geq 1$.
\end{theorem}

Please refer to Appendix~\ref{app:proof-cont-genericity} for details of proof.

\noindent\textbf{Comparative statics.}
As the AI indicator's conditional distributions become more separated in the likelihood ratio sense (e.g., scaling $\Lambda$ by a factor $\kappa > 1$), $G_q$ increases monotonically toward 1, and cascade genericity becomes generic. As $q$ increases (more accurate private impressions), the threshold $\Lq$ rises and $G_q$ falls. The two parameters trade off cleanly: $G_q$ is large when the AI indicator's resolving power dominates the private impression's.

\begin{figure}[t]
  \centering
  \includegraphics[width=0.7\linewidth]{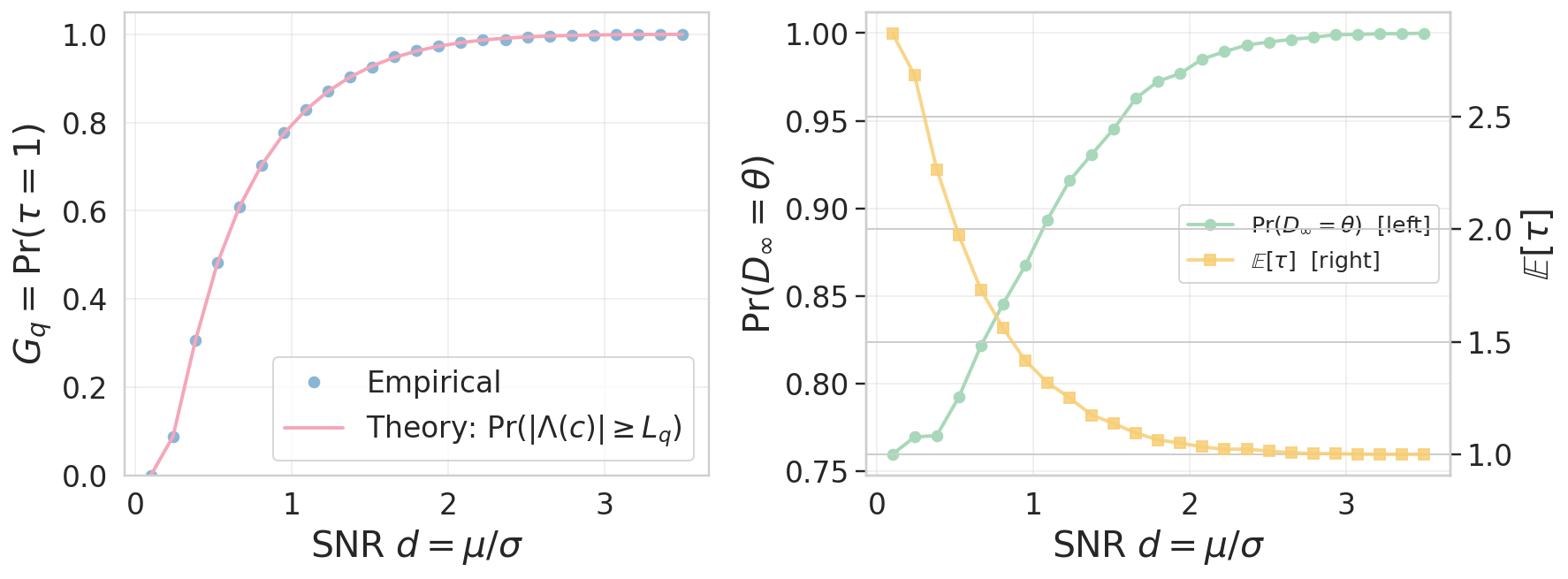}
  \caption{Continuous (Gaussian) extension of the cascade results at $q = 0.7$, $\sigma = 1.0$, $N = 100$. \textbf{Left}: first-user cascade probability $G_q = \Pr(|\Lambda(c)| \geq \Lq)$ as a function of SNR $d = \frac{\mu}{\sigma}$ (Theorem~\ref{thm:cont-genericity}); empirical points match the closed form and rise monotonically to $1$ without discontinuity. \textbf{Right}: long-run consensus accuracy $\Pr(D_\infty = \theta)$ (left axis) and expected cascade time $\E[\tau]$ (right axis) vs SNR.}
  \label{fig:continuous-signal}
\end{figure}

Theorems~\ref{thm:direction} and \ref{thm:spurious} generalize correspondingly. The information aggregation bound becomes $I(\Jbar_N; \theta) \leq I(c; \theta) + I(\sigma_{<\tau}; \theta \mid c)$ with $\sigma_{<\tau}$ defined as in Theorem~\ref{thm:spurious}, the first term collapsing to $I(c; \theta)$ whenever the AI indicator almost surely satisfies the dominance condition $|\Lambda(c)| \geq \Lq$. A continuous AI indicator makes the \emph{first-user} cascade probability $G_q = \Pr(|\Lambda(c)| \geq \Lq)$ smooth in the signal separation, with no analog of the binary discontinuity at $\alpha = q$ (Remark~\ref{rmk:direction-jump}). The conditional learning-region dynamics, however, still inherit the binary-private-signal two-state structure: because the private impression contributes a fixed increment $\pm\Lq$, a starting point $\Lambda(c) \in (0, \Lq)$ visits only the transient states $\{\Lambda(c),\, \Lambda(c) - \Lq\}$, whose transition topology---and hence the gambler's-ruin absorption probabilities $u_+ = \frac{p}{1-p+p^2}$ (from $\Lambda(c) > 0$) and $u_- = \frac{p^2}{1-p+p^2}$ (from $\Lambda(c) < 0$)---do not depend on the value of $\Lambda(c)$. The continuous extension therefore smooths the \emph{frequency} of first-user cascading but preserves, rather than sharpens, the discrete direction-selection content.

\section{Asymmetric prior}\label{app:asymmetric-prior}

The body assumes $\ell_0 = 0$ (symmetric prior). The cascade-genericity threshold extends to $\ell_0 \neq 0$ as follows.

\begin{remark}[Immediate absorption with an asymmetric prior]\label{rmk:asymmetric-prior}
Theorem~\ref{thm:cascade-genericity} is robust to the prior, but the threshold now depends on the realized AI indication $c$. A first-user cascade in the direction of $c$ requires $\alpha \geq \alpha^*(q, \ell_0, c)$, where $\alpha^*(q, \ell_0, c)$ solves $L_{\alpha^*(q, \ell_0, c)} = \Lq - \ell_0 (2c-1)$: a prior aligned with $c$ lowers the threshold, an opposed prior raises it. A single \emph{indication-uniform} condition guaranteeing a first-user cascade for both $c \in \{0,1\}$ is $\La \geq \Lq + |\ell_0|$. In particular, the cascade can be triggered at the first user by an AI indicator of any positive accuracy when the prior is sufficiently informative and aligned with $c$.
\end{remark}

\begin{figure}[t]
  \centering
  \includegraphics[width=0.7\linewidth]{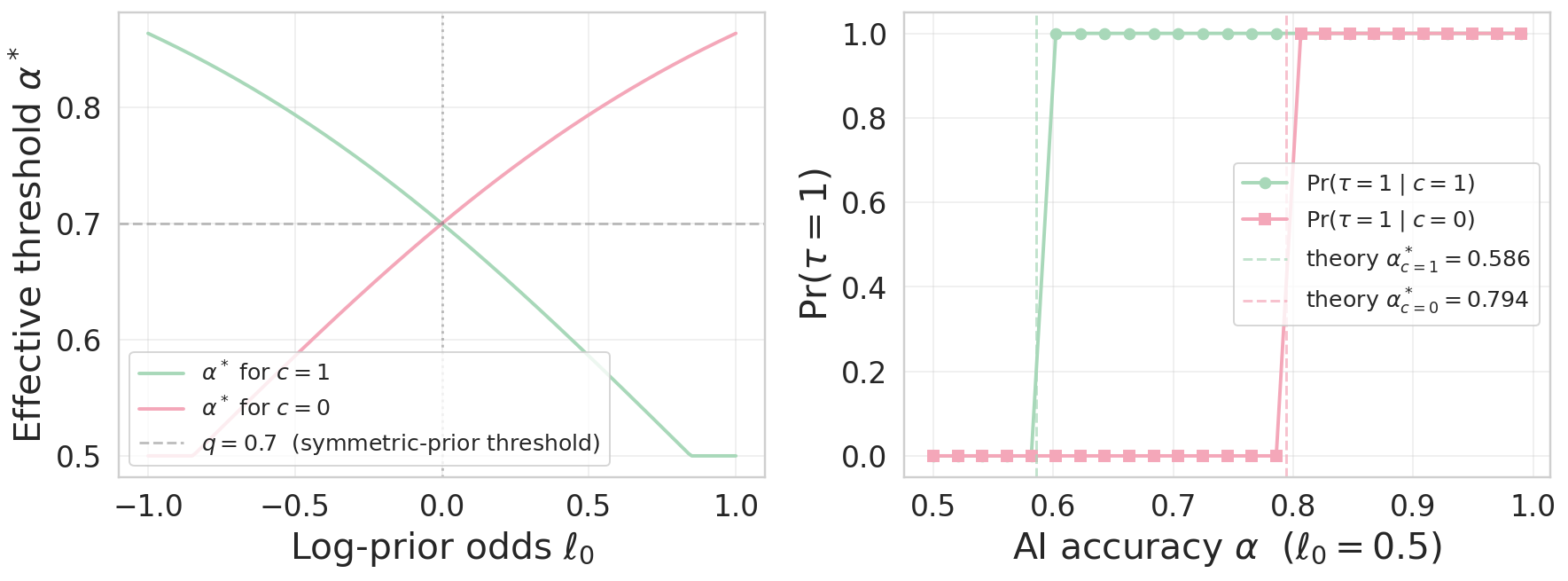}
  \caption{Asymmetric prior at $q = 0.7$. \textbf{Left}: theoretical threshold $\alpha^*(q, \ell_0, c)$ as a function of the prior log-odds $\ell_0$, for each $c \in \{0, 1\}$; a prior aligned with $c$ lowers $\alpha^*$, opposed raises it. \textbf{Right}: empirical first-user cascade probability $\Pr(\tau = 1 \mid c)$ at $\ell_0 = 0.5$, with theoretical thresholds $\alpha^*_{c=1} \approx 0.586$ and $\alpha^*_{c=0} \approx 0.794$ marked as dashed verticals.}
  \label{fig:asymmetric-prior}
\end{figure}

\section{Additional results and figures for the main theorems}\label{app:additional}\label{app:simulations}

This appendix collects the boundary results and supporting figures deferred from Section~\ref{sec:main}: the closed forms for the cascade-formation time, the marginal-monotonicity corollary, the discontinuity remark, and the simulation figures.

\noindent\textbf{Cascade-time closed forms (Proposition~\ref{prop:tau}(c)).}
On the interior $\alpha \in (\frac{1}{2}, q)$, the conditional law of $\tau$ given $(c, \theta)$ depends only on $q$ and on whether $c = \theta$. The first-step probabilities are
\[
\Pr(\tau = 2 \mid c = \theta) = q, \qquad \Pr(\tau = 2 \mid c \neq \theta) = 1 - q,
\]
and the conditional means admit the closed forms
\[
\E[\tau \mid c = \theta] \;=\; 1 + \frac{2-q}{1 - q + q^{2}},
\qquad
\E[\tau \mid c \neq \theta] \;=\; 1 + \frac{1+q}{1 - q + q^{2}},
\]
with difference
\[
\E[\tau \mid c \neq \theta] - \E[\tau \mid c = \theta] \;=\; \frac{2q - 1}{1 - q + q^{2}} \;>\; 0,
\]
so the AI shortens cascade formation when correct and lengthens it when incorrect, the asymmetry governed by $q$ alone.

\begin{corollary}[Marginal monotonicity in $\alpha$]\label{cor:direction-marginal}
For $\alpha \in (\frac{1}{2}, q)$,
\[
\Pr(D_\infty = \sign(\lambda(c;\alpha)) \mid c)
\;=\;
\frac{(1-q) + \alpha(2q - 1)}{1 - q + q^{2}},
\]
which is strictly increasing in $\alpha$. The monotonicity is sourced entirely from the posterior $\Pr(\theta = c \mid c) = \alpha$; the chain dynamics on $\{A, B\}$ contribute no $\alpha$-dependence. (Proof in Appendix~\ref{app:proof-direction-marginal}.)
\end{corollary}

\begin{remark}[Discontinuity at $\alpha = q$]\label{rmk:direction-jump}
The map $\alpha \mapsto \Pr(D_\infty = \sign(\lambda(c;\alpha)) \mid c)$ is discontinuous at $\alpha = q$. On $(\frac{1}{2}, q)$ it approaches
\[
\lim_{\alpha \to q^{-}} \Pr(D_\infty = \sign(\lambda(c;\alpha)) \mid c) = \frac{q^{2} + (1-q)^{2}}{1 - q + q^{2}} = \frac{1 - 2q(1-q)}{1 - q + q^{2}} \;<\; 1,
\]
and jumps to $1$ at $\alpha = q$ because Theorem~\ref{thm:cascade-genericity} forces a first-user cascade as soon as $\La \geq \Lq$. The interior limit governs only $(\frac{1}{2}, q)$ and stays strictly below $1$; on regime~(a) ($\alpha \geq q$) the probability is identically $1$, so the value at $\alpha = q$ is reached by the jump, not as a limit from within the interior.
\end{remark}

\begin{figure}[t]
  \centering
  \includegraphics[width=0.7\linewidth]{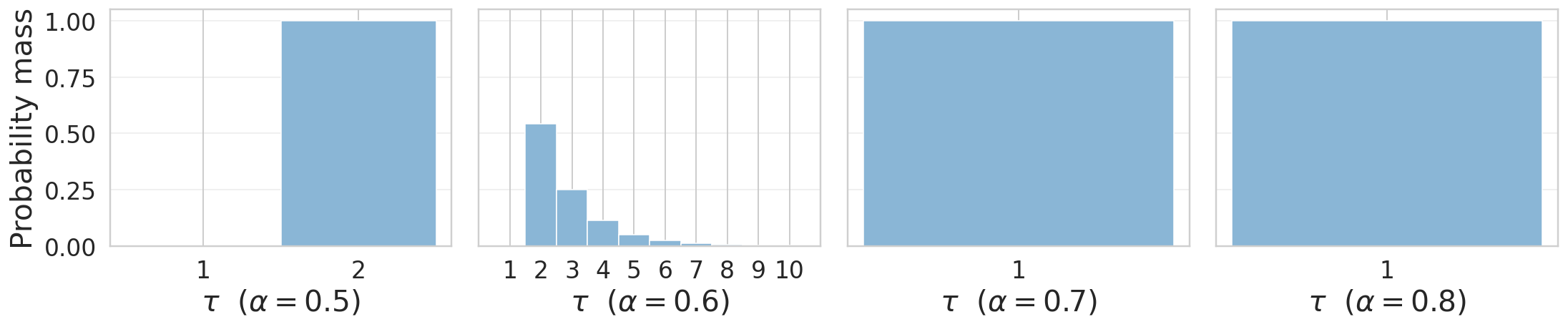}
  \caption{Distribution of cascade-formation time $\tau$ at private accuracy $q = 0.7$ over $20{,}000$ simulated cascades of $N = 50$ users (Theorem~\ref{thm:cascade-genericity}). Each panel is a different AI accuracy $\alpha$ (annotated on the panel's $x$-axis). The probability mass at $\tau = 1$ rises through the learning region $\alpha \in (\tfrac{1}{2}, q)$ and jumps to $1$ at $\alpha = q$.}
  \label{fig:tau-by-alpha}
\end{figure}

\begin{figure}[t]
  \centering
  \includegraphics[width=0.7\linewidth]{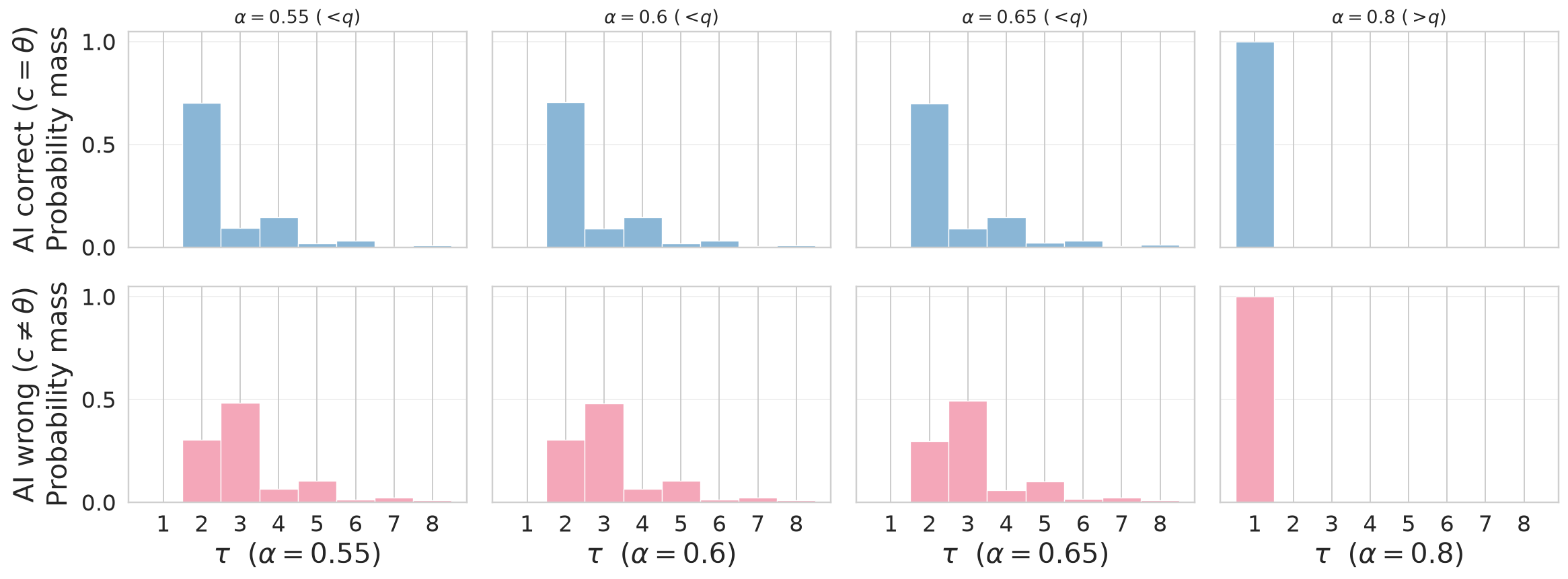}
  \caption{Distribution of $\tau$ on the interior $\alpha \in (\tfrac{1}{2}, q)$ at $q = 0.7$, split by AI correctness (Proposition~\ref{prop:tau}(c)). \textbf{Top row} (blue): $c = \theta$, mass concentrates at small $\tau$. \textbf{Bottom row} (pink): $c \neq \theta$, distribution shifts to larger $\tau$. Each column is a different $\alpha$, annotated on the bottom $x$-axis.}
  \label{fig:tau-split}
\end{figure}

\begin{figure}[t]
  \centering
  \includegraphics[width=0.7\linewidth]{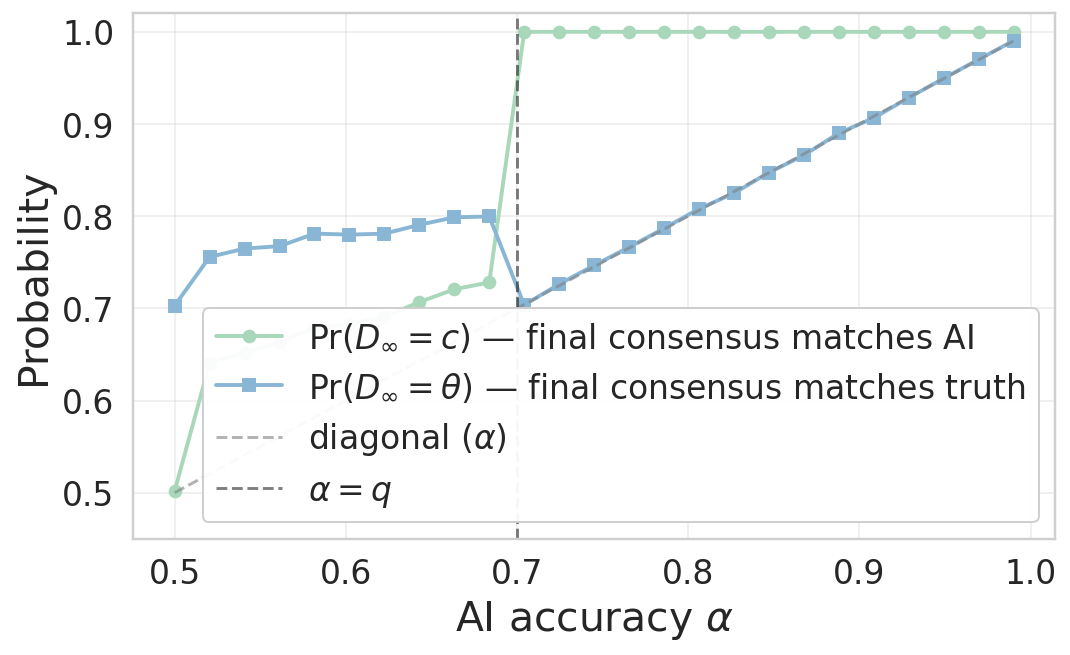}
  \caption{Marginal $\Pr(D_\infty = \sign(\lambda(c;\alpha)) \mid c)$ as a function of $\alpha$ at $q = 0.7$ (Corollary~\ref{cor:direction-marginal}). The curve rises monotonically on $(\tfrac{1}{2}, q)$ and jumps to $1$ at $\alpha = q$ (Remark~\ref{rmk:direction-jump}).}
  \label{fig:direction-selection}
\end{figure}

\begin{figure}[t]
  \centering
  \includegraphics[width=0.7\linewidth]{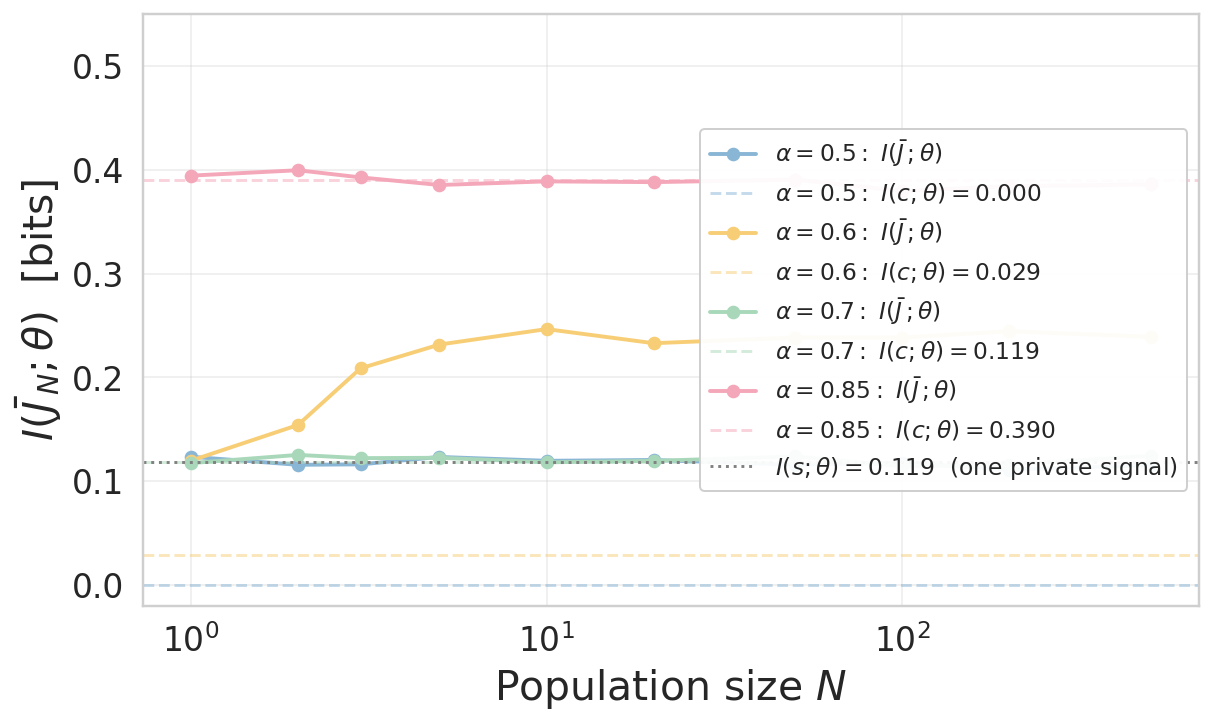}
  \caption{Mutual information $I(\Jbar_N; \theta)$ versus population size $N$ for $\alpha \in \{\tfrac{1}{2}, 0.6, 0.75\}$ at $q = 0.7$ (Theorem~\ref{thm:spurious}). The curves saturate at finite ceilings independent of $N$; the dashed line marks the strong-AI tight value $I(c; \theta)$.}
  \label{fig:info-saturation}
\end{figure}

\begin{figure}[t]
  \centering
  \includegraphics[width=0.7\linewidth]{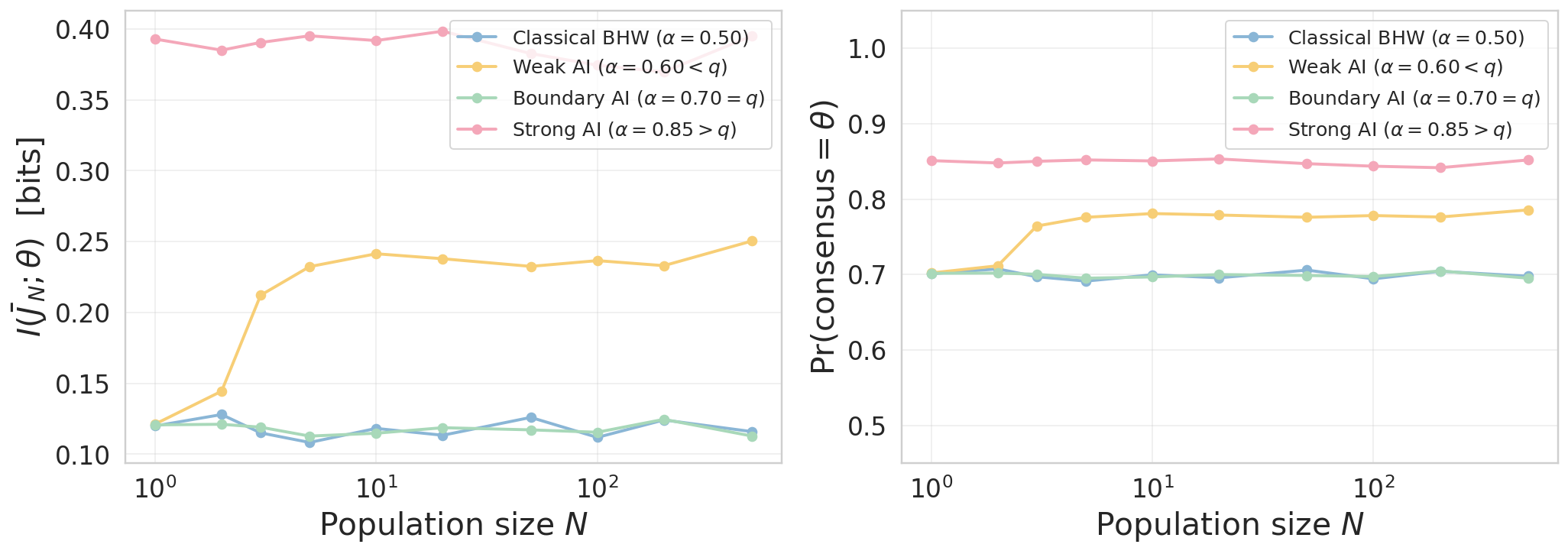}
  \caption{Classical BHW vs.\ the augmented model at $q = 0.7$, with $\alpha \in \{0.50, 0.60, 0.70, 0.85\}$ spanning the classical limit, learning regime, boundary, and strong-AI regime---an extended view of Figure~\ref{fig:info-saturation}. \textbf{Left}: mutual information $I(\bar J_N; \theta)$ vs population size $N$ saturates in every regime---users arriving after cascade formation add no independent evidence. \textbf{Right}: long-run consensus accuracy $\Pr(\mathrm{consensus} = \theta)$ vs $N$ plateaus quickly, with the initial regime alone determining the ceiling.}
  \label{fig:classical-vs-augmented}
\end{figure}

\section{Dataset and experimental design}\label{app:data}

We use the experimental data from a prior work of AI-based credibility indicators in social misinformation cascades. Each room is one cascade: a single news item evaluated sequentially by eleven subjects drawn from a pool of several hundred recruited participants. Subjects rotate across rooms (the same subject may occupy different positions in different rooms), so the model's in-room assumption of sequential, independent private signals is preserved, but rooms are not independent across the corpus---a limitation we return to in Section~\ref{sec:discussion}. The corpus comprises $538$ distinct subjects contributing about ten judgments each, with no subject spanning more than one treatment (the design is between-subjects); it is this repeated-measures structure---each subject seen under varying AI signals, public tallies, and private impressions---that identifies the per-subject behavioral weights of Section~\ref{sec:emp-behavioral}. Each record indexes one room by $(group, news\_index, room\_id)$, with $480$ rooms total ($160$ per treatment, $40$ news items per treatment, $4$ rooms per $(group, news\_index)$); the item-level analysis of Appendix~\ref{app:emp-diagnostics} pools the two AI treatments, so each of the $40$ items is observed in $8$ rooms.

Each room is one realization of the model in Section~\ref{sec:model} with $N = 11$ users. The state $\theta$ is the news item's veracity, fixed and known to the experimenter; the AI signal $c$ is a binary credibility prediction generated once per news item by a held-out classifier and held fixed within a room; each subject publishes a binary judgment $J_i \in \{0, 1\}$ (fake/real), which enters the public history visible to subsequent subjects in the same room. We work directly with these per-user judgments, recoded to $\{-1, +1\}$ for the log-odds quantities. Each subject also records a \emph{pre-crowd initial judgment}, formed before seeing the public history (and, in \textsc{AI-after}, before the AI signal); this provides a direct measure of the private signal $s_i$, from which we identify $\widehat q$.

Three between-subjects treatments vary the presence and timing of the AI-based credibility indicator:
\begin{description}[leftmargin=2em]
\item[\textsc{Control}] no AI indicator is shown; this implements the classical BHW limit $\alpha = \frac{1}{2}$.
\item[\textsc{AI-before}] the AI signal is shown in Step 1, before the subject's own initial impression and before the public history; this matches the canonical information set $(s_i, c, H_i)$ of Section~\ref{sec:model}.
\item[\textsc{AI-after}] the AI signal is shown in Step 3, alongside the public history, after the subject has formed and recorded an initial private impression; the subject's published judgment $J_i$ is then formed conditional on $(s_i, c, H_i)$ as in the model, but the recorded initial impression provides a separately observable measure of the private signal $s_i$.
\end{description}
The three treatments are balanced ($160$ rooms each) and share the same set of $40$ news items, so $\alpha$ is held fixed across treatments by design.

\section{Full empirical diagnostics}\label{app:emp-diagnostics}

This appendix expands the calibration of Section~\ref{sec:empirical}: the four per-prediction tests of the strict benchmark, the behavioral model's specification and identification, and the item-level residual the per-user fit does not capture.

\subsection{Strict-benchmark diagnostics}

\begin{figure}[t]
  \centering
  \includegraphics[width=0.6\linewidth]{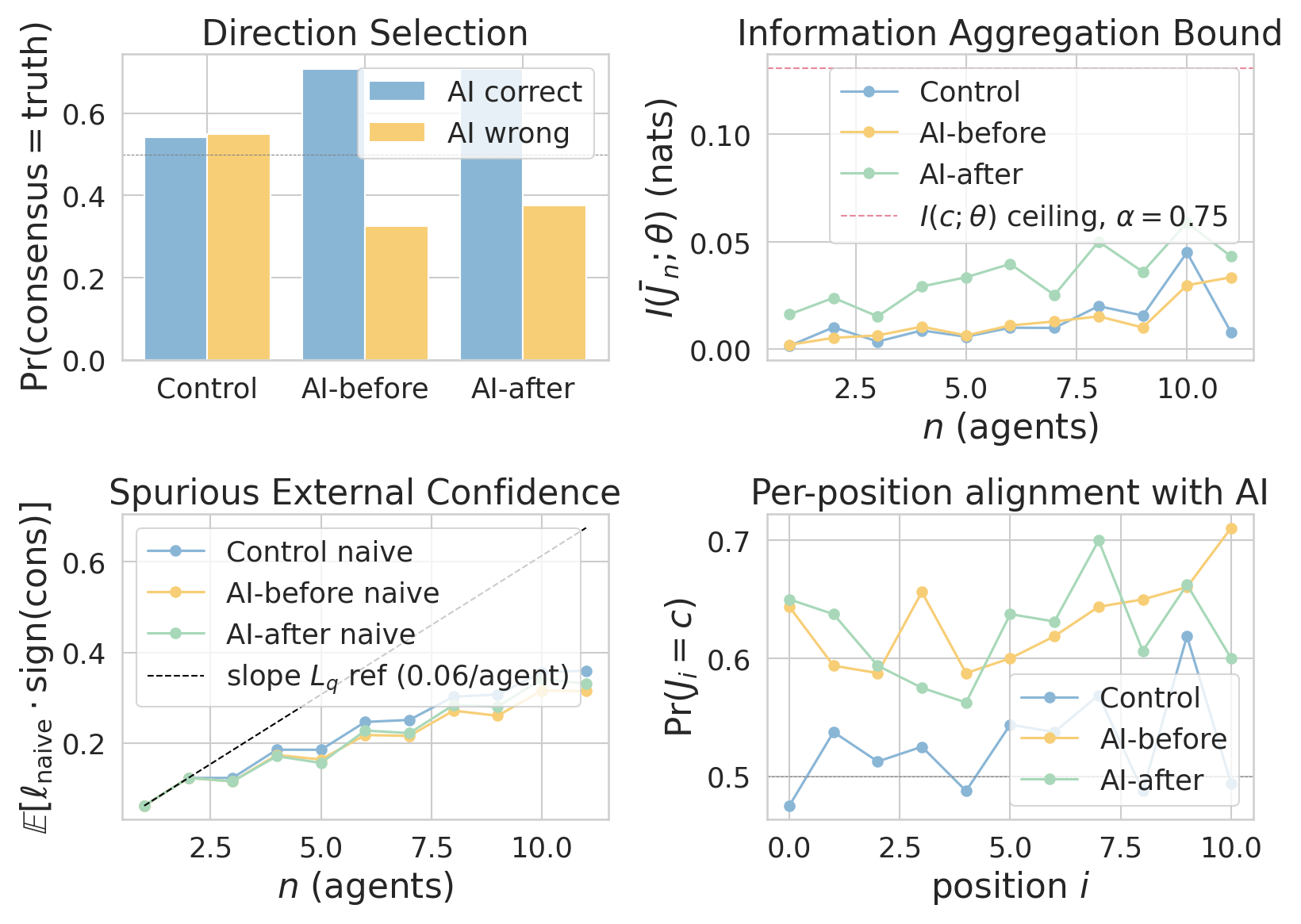}
  \caption{Strict-benchmark diagnostics on the experimental dataset ($480$ rooms, $N = 11$ subjects per room, three treatments). \textbf{(a)} Direction selection: $\Pr(\mathrm{consensus} = \theta)$ by group, split by AI correctness. \textbf{(b)} Information aggregation: plug-in $\widehat I(\Jbar_n; \theta)$ versus $n$ by group, with the strong-AI ceiling $I(c;\theta)$ at $\widehat\alpha = 0.75$ marked. \textbf{(c)} Spurious external confidence: running mean of $\ell_{\mathrm{ext}}^{\mathrm{naive}} \cdot \sign(\mathrm{consensus})$ versus $n$, with the theoretical slope $\widehat\Lq$ as reference. \textbf{(d)} Per-position $\Pr(J_i = c)$ across positions $i = 1, \ldots, 11$.}
  \label{fig:empirical-validation}
\end{figure}

Because $\widehat\alpha > \widehat q$, the strict model lives in the strong-AI regime, where its predictions are sharp and parameter-free. We test all four in turn on the per-user judgments; the diagnostics are visualized in Figure~\ref{fig:empirical-validation} (Section~\ref{sec:empirical}).

\noindent\textbf{First-user cascade (Theorem~\ref{thm:cascade-genericity}).}
The strict prediction is $\Pr(J_i = c) = 1$ at every position $i$ in both AI treatments. The empirical first-user rate is $0.644$ (\textsc{AI-before}) and $0.650$ (\textsc{AI-after}), and the per-position rate $\Pr(J_i = c)$ stays in the band $0.56$--$0.71$ across all eleven positions with no tendency toward $1$. The fraction of rooms in which all eleven judgments match $c$ is only $0.044$ (\textsc{AI-before}) and $0.119$ (\textsc{AI-after}), not the $\approx 1$ implied by strict cascade locking. The point prediction is \emph{rejected}.

\noindent\textbf{Direction selection (Theorem~\ref{thm:direction}).}
Regime (a) predicts $\Pr(\mathrm{majority} = c) = 1$, i.e.\ majority tracks truth with probability $1$ when the AI is correct and $0$ when wrong. Conditional on a correct AI, the majority matches the truth in $0.708$ (\textsc{AI-before}) and $0.708$ (\textsc{AI-after}) of rooms; conditional on a wrong AI, in $0.325$ and $0.375$. The strict $0/1$ prediction is \emph{rejected}. The qualitative comparative content survives: when the AI is wrong, both AI treatments fall well below the $50\%$ truth-accuracy line, while \textsc{Control} stays roughly symmetric around $0.54$, consistent with the no-AI baseline ($\alpha=\tfrac{1}{2}$).

\noindent\textbf{Information aggregation (Theorem~\ref{thm:spurious}).}
The plug-in $\widehat I(\Jbar_n; \theta)$ is computed from the empirical joint histogram of $(\sign(\Jbar_n), \theta)$ over the $480$ rooms. \textsc{Control} stays near zero ($\leq 0.045$ nats), \textsc{AI-before} grows modestly ($\leq 0.034$), and \textsc{AI-after} rises to a plateau of $0.04$--$0.06$. The strong-AI tight value $I(c;\theta) = 0.131$ nats at $\widehat\alpha = 0.75$ is \emph{not attained}; the realized plateau is well below it. But the qualitative content is consistent with a finite, $N$-independent ceiling: across the eleven positions none of the three curves shows evidence of the monotone growth in $n$ that wisdom-of-crowds aggregation would predict. (We read this as consistent with, not a test of, the asymptotic statement, given the short $N = 11$ horizon.)

\noindent\textbf{Spurious external confidence (Theorem~\ref{thm:spurious}).}
The slope is where the benchmark comes closest to a quantitative match. The running mean of $\ell_{\mathrm{ext}}^{\mathrm{naive}} \cdot \sign(\mathrm{consensus})$ grows near-linearly in $n$, with OLS slope $\approx 0.03$ per user across all three groups---the same order as $\widehat\Lq = 0.061$, the small shortfall reflecting imperfect cascade alignment. The amplitude constant of Theorem~\ref{thm:spurious} is untestable here: the naive--correct crossover is at $N^\dagger \approx \frac{\widehat\La}{\widehat\Lq} \approx 18$, beyond the $N = 11$ horizon. Only the slope, the regime-independent part of the prediction, is in range, and it holds.

\subsection{The behavioral model}

We keep the model's information structure (the public-history component, the private contribution, and the AI indicator) but relax the strict Bayesian \emph{decision rule} with a few free parameters. We use a reduced-form logit in the quantal-response / behavioral-aggregation tradition; it nests three relaxations:
\begin{enumerate}
\item[(i)] \textbf{Softmax choice} in place of the hard threshold: $\Pr(J_i = +1) = \sigma(\nu \cdot \ell_i)$, with temperature $\nu$ capturing decision noise; $\nu \to \infty$ recovers the deterministic rule of Section~\ref{sec:model}.
\item[(ii)] \textbf{Weighted (not equal) aggregation} of the public-history and private channels, which enter the index with separate weights rather than at the common unit weight the strict rule imposes.
\item[(iii)] \textbf{Over/under-weighting of the AI indicator} through a separate weight on $c$, capturing relative over- or under-weighting of the AI log-odds channel against the public-history channel.
\end{enumerate}
We estimate the nested specification
\[
\mathrm{logit}\,\Pr(J_i = +1) \;=\; b_0
  \;+\; w_{\mathrm{public}}\, \big(\Lq\, T_i\big)
  \;+\; w_{\mathrm{AI}}\, \big(\La\, c_i\big)
  \;+\; w_{\mathrm{private}}\, \theta_i ,
\]
on all judgments, where $T_i = \sum_{k < i} J_k$ is the signed public tally, $c_i = 2c-1$ is the AI signal, and $\theta_i = 2\theta - 1$ is the truth. The public and AI regressors are placed in \emph{log-odds units} ($\Lq T_i$ is the naive public log-odds, $\La c_i$ the AI log-odds), so $w_{\mathrm{public}}, w_{\mathrm{AI}}$ are dimensionless multipliers on log-odds channels. We use $\theta_i$ as a \emph{noisy proxy} for the unobserved private impression $s_i$ (subjects observe $s_i$, not $\theta$); we therefore read $w_{\mathrm{private}}$ as a \emph{residual truth-aligned sensitivity} after public and AI controls, not as a clean private-channel weight.

\noindent\textbf{Identification.} Under a logit/QRE link the index is $\nu \cdot \ell_i$, so each fitted coefficient equals $\nu$ times a structural weight: the overall scale is the (unidentified) temperature $\nu$, and only coefficient \emph{ratios} are identified without a normalization. We therefore take the temperature-free per-signal ratio $\rho_{\text{AI-public}} = \frac{w_{\mathrm{AI}}\,\widehat\La}{w_{\mathrm{public}}\,\widehat\Lq}$---the contribution of one AI indication relative to one public judgment, invariant to the log-odds scaling of the regressors---as the robust object, and report any absolute $\widehat\beta_{\mathrm{AI}}$ only under an explicitly stated normalization. Standard errors in Table~\ref{tab:behavioral-fit} are two-way clustered by room and news item (Cameron--Gelbach--Miller); they are nearly identical under room-only or item-only clustering, so the strong item-level correlation documented below does not overturn the significance pattern. $w_{\mathrm{private}}$ is the only coefficient whose $95\%$ CI contains zero in the AI groups.

\noindent\textbf{(i) Public history and the AI indicator contribute comparably.}
The sequential ablation is clear: adding the public tally raises the log-likelihood by $+36$ (\textsc{AI-before}) to $+54$ (\textsc{AI-after}), and adding the AI indicator by a further $+42$ and $+32$, while the truth proxy adds essentially nothing. Both channels are tightly estimated ($w_{\mathrm{public}} \approx 1.4$--$2.2$ and $w_{\mathrm{AI}} \approx 0.42$--$0.47$, clustered $t > 4$); neither dominates. That subjects respond to the accumulated (cascade-correlated) public record as if it were strong evidence is the within-user counterpart of the naive external observer of Theorem~\ref{thm:spurious}---though the regression alone cannot separate belief updating from conformity.

\noindent\textbf{(ii) Residual truth sensitivity is small.}
After public and AI controls, $w_{\mathrm{private}}$ is small and not significantly different from zero in either AI group ($0.01 \pm 0.06$, $-0.01 \pm 0.06$, two-way clustered; adding it leaves $\log L$ essentially unchanged). Because $\theta$ is only a proxy for $s_i$, we do not interpret this as ``the private channel is silent'' in a structural sense; we read it as the data showing no detectable truth-aligned pull beyond what the public and AI channels already carry---consistent with $\widehat q \approx \tfrac{1}{2}$, under which a private impression conveys little.

\noindent\textbf{(iii) One AI prediction carries the weight of several public judgments.}
The temperature-free per-signal ratio is $\rho_{\text{AI-public}} \approx 5.9$ (\textsc{AI-before}) and $3.5$ (\textsc{AI-after}), matching the EM reduced-form estimates ($6.11$ and $3.52$, Appendix~\ref{app:em}): one AI prediction contributes as much to the fitted choice index as roughly three to six public-history judgments. Equivalently, per log-odds unit the AI is weighted at $\rho_{\text{AI-public}} \cdot \frac{\widehat\Lq}{\widehat\La} \approx 0.20$--$0.33$ of the public-history multiplier---a moderate algorithm aversion relative to the social channel---though this second reading, unlike the per-signal ratio, depends on the point value of $\widehat\Lq$. The \emph{direction} of the gap is robust because it is a ratio: it identifies under-weighting of the AI \emph{relative to the public-history channel}, not relative to the Bayes-optimal AI weight, which would require pinning down the temperature normalization.

The \emph{absolute} discount, by contrast, is normalization-dependent. Writing $w_{\mathrm{AI}} = \nu \cdot \beta_{\mathrm{AI}}$, the implied $\widehat\beta_{\mathrm{AI}}$ is $\approx 0.42$--$0.47$ under the normalization $\nu = 1$ (logit scale equals the Bayesian log-odds scale) and $\approx 0.20$--$0.33$ under the normalization $w_{\mathrm{public}} = 1$ (public history taken at face value). We therefore decline to report a single structural $\beta_{\mathrm{AI}}$ without committing to a temperature normalization; the robust, normalization-free statement is that one AI prediction carries the weight of several public judgments, even as per log-odds unit the AI is weighted moderately below the social channel.

\begin{table}[t]
\centering
\small
\begin{tabular}{lccc}
\toprule
 & \textsc{AI-before} & \textsc{AI-after} & \textsc{Control} \\
\midrule
\multicolumn{4}{l}{\emph{Full model (two-way clustered SE by room and item)}}\\
$w_{\mathrm{public}}$ (public history) & $1.41\;(0.32)$ & $2.18\;(0.27)$ & $2.15\;(0.37)$ \\
$w_{\mathrm{AI}}$ (AI log-odds)     & $0.47\;(0.06)$ & $0.42\;(0.06)$ & --- \\
$w_{\mathrm{private}}$ (truth proxy)   & $0.01\;(0.06)$ & $-0.01\;(0.06)$ & $0.18\;(0.07)$ \\
$\rho_{\text{AI-public}} = \frac{w_{\mathrm{AI}}\,\widehat\La}{w_{\mathrm{public}}\,\widehat\Lq}$ & $5.93$ & $3.51$ & --- \\
\midrule
\multicolumn{4}{l}{\emph{Sequential ablation: incremental $\Delta\log L$ from}}\\
\multicolumn{4}{l}{\emph{adding public history, then AI, then truth proxy}}\\
$+\,$public history & $+36.4$ & $+54.0$ & $+29.7$ \\
$+\,$AI indicator   & $+42.4$ & $+32.1$ & --- \\
$+\,$truth proxy    & $+0.0$  & $+0.0$  & $+5.6$ \\
\midrule
$n$ (judgments) / rooms & $1758/160$ & $1760/160$ & $1760/160$ \\
\bottomrule
\end{tabular}
\caption{Pooled behavioral decision-rule fit (the single-class baseline of the
per-subject mixture of Section~\ref{sec:mixture}). Coefficients are
dimensionless multipliers on log-odds channels; the absolute scale is the
(unidentified) temperature, so the temperature-free per-signal ratio
$\rho_{\text{AI-public}}$ is the robust quantity, directly comparable with
the EM reduced-form estimates ($6.11$ and $3.52$, Appendix~\ref{app:em}). SEs (in parentheses) are two-way clustered by room and news
item. Full specification, identification, and robustness in
Appendix~\ref{app:em}.}
\label{tab:behavioral-fit}
\end{table}

\subsection{Per-subject behavioral weights via EM}\label{app:em}

Section~\ref{sec:emp-behavioral} places the population on the regime map using a temperature-free weight ratio backed out per subject. We give the specification here. For each published judgment we form the index regressors $c = 2c-1$ (the AI signal), $T_i = \sum_{k<i} J_k$ (the signed public tally before position $i$), and $s = 2s_i - 1$ (the private impression, read off the pre-crowd initial judgment), and model
\[
\mathrm{logit}\,\Pr(J_i = +1) \;=\; b_0 + \beta_{\mathrm{AI}}\,c + \beta_{\mathrm{public}}\,T_i + \beta_{\mathrm{private}}\,s,
\]
a softmax/QRE decision rule whose temperature is absorbed into $(\beta_{\mathrm{AI}},\beta_{\mathrm{public}},\beta_{\mathrm{private}})$. The treatment determines which regressors are clean. In \textsc{AI-after} the initial judgment is formed before the AI and the crowd, so $s$ is a clean private impression and we fit the full $1+c+T+s$; this is the only group in which the cascade-threshold ratio $\rho_{\text{AI-private}}$ is identifiable. In \textsc{Control} there is no AI, so we fit the classical $1+T+s$ and read $\rho_{\text{public-private}}$. In \textsc{AI-before} the initial judgment is formed \emph{after} the AI and is therefore a mediator of $c$ rather than a clean private impression; including it would leak the AI effect into $\beta_{\mathrm{private}}$, so we drop it and fit the reduced form $1+c+T$, which identifies the AI-versus-public ratio $\rho_{\text{AI-public}}$ but not $\rho_{\text{AI-private}}$.

Because subjects recur across rooms, we fit a mixture of these logits with the subject as the grouping unit: each subject $w$ has a latent class $z_w \in \{1,\dots,K\}$, and all of $w$'s judgments share the class-$k$ coefficients $\beta_k$. EM alternates an E-step, in which each subject's responsibility $\gamma_{wk}\propto\pi_k\prod_{j\in w}\Pr(J_j\mid x_j,\beta_k)$ is the product of her judgment likelihoods under class $k$, and an M-step, in which each class is refit by responsibility-weighted logistic regression; we select $K\in\{1,2,3\}$ by BIC with the subject count as the sample size. The pooled fit of Table~\ref{tab:behavioral-fit} is the $K=1$ limit.

\noindent\textbf{Identification.} Only ratios are identified: the index is $\nu\cdot\ell$ for an unidentified temperature $\nu$, so the fitted $\beta$'s are each $\nu$ times a structural weight and only the ratios $\rho_{\text{AI-private}} \equiv \frac{\beta_{\mathrm{AI}}}{\beta_{\mathrm{private}}}$, $\rho_{\text{AI-public}} \equiv \frac{\beta_{\mathrm{AI}}}{\beta_{\mathrm{public}}}$, $\rho_{\text{public-private}} \equiv \frac{\beta_{\mathrm{public}}}{\beta_{\mathrm{private}}}$ are temperature-free. We report subject-resampling bootstrap CIs (resample subjects with replacement, refit the pooled model, $B=400$).

\noindent\textbf{Results.} In \textsc{AI-after} the pooled cascade-threshold ratio is $\rho_{\text{AI-private}} = 0.49$ (CI $[0.33,0.67]$, below $1$ in every resample), with $\rho_{\text{AI-public}} = 3.35$; in \textsc{Control} $\rho_{\text{public-private}} = 0.12$ (CI $[0.08,0.16]$) against $0.15$ in \textsc{AI-after}, so the AI does not disturb the social/private balance. The BIC-selected $K=3$ mixture splits \textsc{AI-after} subjects into a learner majority ($\approx 46\%$, $\rho_{\text{AI-private}}\approx 0.3$), a cascade-prone component ($\approx 20\%$, $\rho_{\text{AI-private}}\approx 1.8$), and a weakly-identified remainder ($\approx 34\%$); \textsc{Control} has no class crossing the crowd-cascade threshold $\rho_{\text{public-private}}\ge 1$.

\noindent\textbf{Timing, apples-to-apples.} The AI-before-versus-AI-after gap is not an artefact of dropping $s$: under the identical reduced-form $1+c+T$ for both groups, $\rho_{\text{AI-public}} = 6.11$ (\textsc{AI-before}) against $3.52$ (\textsc{AI-after}). Re-including $s$ in \textsc{AI-after} barely moves its ratio ($\rho_{\text{AI-public}}: 3.52 \to 3.35$), as expected when $\widehat q\approx\tfrac12$ makes the private impression nearly uninformative, so the timing contrast is genuine: showing the AI before the subject forms her own impression anchors her judgment harder than showing it after.

\subsection{Item-level clustering: a residual outside the per-user model}

The fit captures \emph{individual decision behavior} under a stated normalization---the relative public/AI/truth weights---which is what Proposition~\ref{prop:behavioral} concerns and what a reduced-form test can identify. It does not capture one feature of the data that is \emph{cross-sectional} rather than per-user: the herd direction clusters strongly by news item (the per-item rate of agreement with $c$ ranges from $0.12$ to $1.00$ across the $40$ items, with between-item over-dispersion $\approx 3.3\times$ the within-item binomial benchmark), and about a third of rooms cascade \emph{against} $c$ independently of AI correctness. Per-user parameters cannot generate item-level clustering; it points to a shared, item-level component of impressions beyond the i.i.d.\ private-impression assumption of Section~\ref{sec:model}---a correlated-signal extension (which would relax the conditional independence that yields the scalar chain $(P_i)$) that we leave to future work and flag in Section~\ref{sec:discussion}.

\section{Behavioral robustness: setup, comparative statics, and simulations}\label{app:behavioral-full}

This appendix expands Section~\ref{sec:emp-behavioral}.

\noindent\textbf{Behavioral public log-odds.}
In $\ell_i^\beta$, the term $\psi_j(J_j)$ is the log-LR that $J_j$ carries about $\theta$ as perceived by later users applying the same behavioral rule, not the Bayesian one. Collecting everything user $i$ sees except her own impression gives the behavioral public log-odds
\[
P_i^\beta \;\equiv\; \ell_0 + \beta_{\mathrm{AI}} \cdot \lambda(c; \alpha) + \beta_{\mathrm{public}} \cdot \sum_{j < i} \psi_j(J_j),
\]
so that $\ell_i^\beta = P_i^\beta + \beta_{\mathrm{private}} \cdot \lambda(s_i; q)$. Proposition~\ref{prop:behavioral} uses only the case in which forced judgments contribute $\psi_j = 0$, so its conclusion does not depend on the full behavioral $\psi$-dynamics.

\noindent\textbf{Comparative statics on $P_1$.}
(a) AI-indicator aversion ($\beta_{\mathrm{AI}} < 1$) raises $\alpha^{*}$ and weakens first-user cascade genericity; AI-indicator appreciation ($\beta_{\mathrm{AI}} > 1$) lowers $\alpha^{*}$. (b) Conservatism ($\beta_{\mathrm{private}} < 1$) lowers $\alpha^{*}$, making first-user cascade easier to trigger. (c) Social discounting ($\beta_{\mathrm{public}} < 1$) does not enter the first-user threshold (no public history at $i = 1$) but locally weakens the social term, which should tend to enlarge the learning region of Proposition~\ref{prop:tau} and delay absorption when the chain does not cascade at user 1. We do not claim a closed-form analog of Proposition~\ref{prop:tau} under $\beta_{\mathrm{public}} \neq 1$; only the qualitative direction follows from the comparative statics.

\begin{figure}[t]
  \centering
  \includegraphics[width=0.7\linewidth]{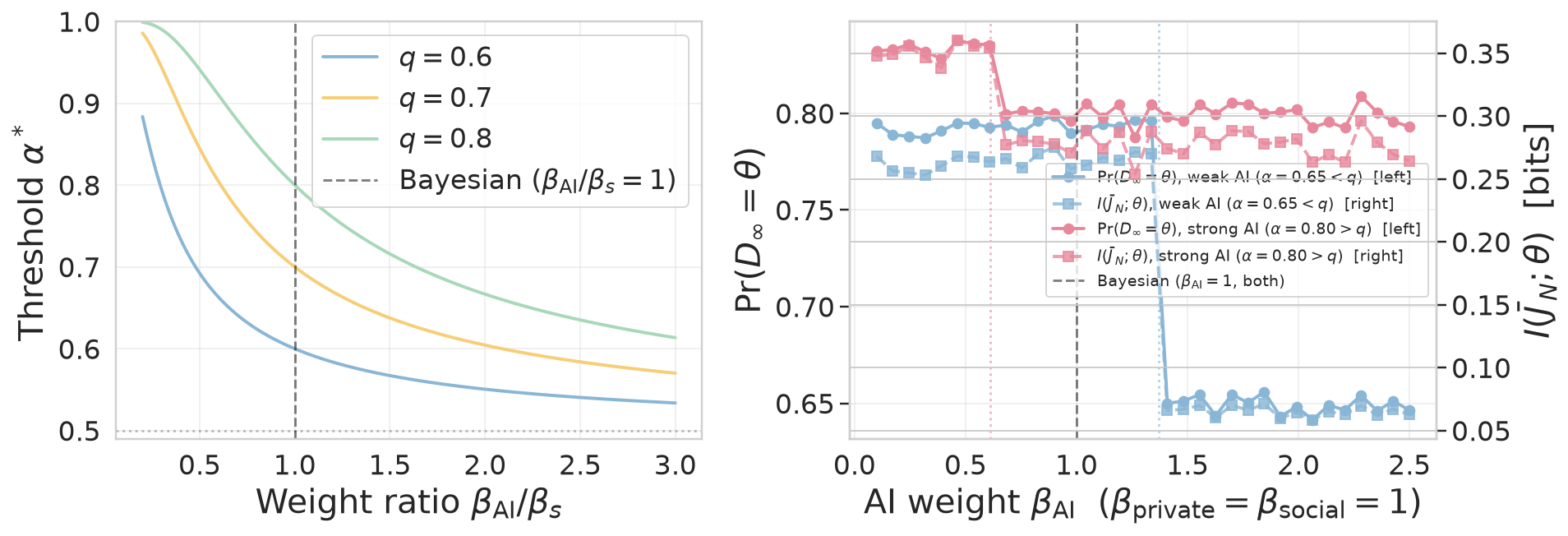}
  \caption{Behavioral robustness of the first-user threshold (Proposition~\ref{prop:behavioral}). \textbf{Left}: effective cascade-genericity threshold $\alpha^*$ as a function of the weight ratio $\frac{\beta_{\mathrm{AI}}}{\beta_{\mathrm{private}}}$ for $q \in \{0.6, 0.7, 0.8\}$; AI-aversion ($\frac{\beta_{\mathrm{AI}}}{\beta_{\mathrm{private}}} < 1$) raises $\alpha^*$, AI-appreciation lowers it. \textbf{Right}: simulation under the weighted decision rule at $\alpha = 0.65 < q = 0.7$ and $N = 100$, where AI-appreciation ($\beta_{\mathrm{AI}}$ above the behavioral threshold $\beta^* = \frac{L_q}{L_\alpha}$) flips the chain into immediate cascade and collapses private-signal aggregation; $\Pr(D_\infty = \theta)$ on the left axis, $I(\bar J_N; \theta)$ on the right.}
  \label{fig:behavioral-robustness}
\end{figure}

\section{Scope of the behavioral threshold proposition}\label{app:behavioral-scope}

Proposition~\ref{prop:behavioral} is deliberately limited to the \emph{first-user threshold} and its absorbing consequence. We do not claim that Theorems~\ref{thm:direction} and~\ref{thm:spurious} extend verbatim to the behavioral model. The reason is that those theorems rely on Bayesian users using Lemma~\ref{lem:cascade-structure}'s ``$\psi_j = \pm\Lq$ or 0'' increment structure; under heterogeneous or non-unit $\beta$'s, downstream users' inference about prior users' private impressions from observed judgments is no longer the clean $\pm\Lq$ jump, and Proposition~\ref{prop:tau}'s two-state $\{A, B\}$ reduction need not hold. Extending the direction-selection, information-bound, and spurious-confidence theorems to a behavioral population is a non-trivial open problem; we record it as such and offer the present proposition only as a comparative-statics statement about the first-user threshold.

\section{Delayed exposure: when to attach the indicator}\label{app:delay}\label{sec:delay}\label{sec:emp-deploy}

Beyond how strongly the indicator is trusted (Section~\ref{sec:perceived}), a platform also controls \emph{when} it is attached. This deployment lever is not structural---it changes neither the model nor the user---and we treat it here in simulation. Like miscalibrated trust, it does not escape the welfare frontier of Proposition~\ref{prop:frontier}; it fails in a different way, by separating the recorded majority from the direction the crowd actually takes.

\begin{figure}[t]
  \centering
  \includegraphics[width=\linewidth]{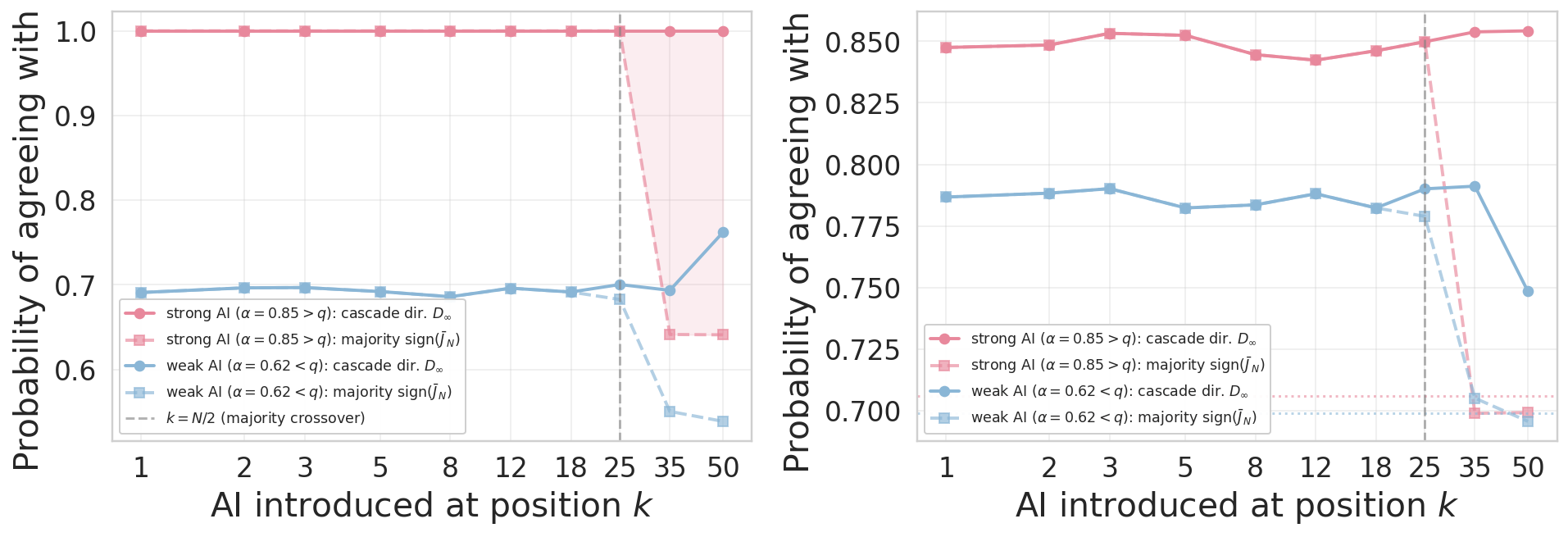}
  \caption{Delayed exposure: the indicator is attached from position $k$, with users $1, \dots, k-1$ in the classical model ($q = 0.7$, $N = 50$, $12{,}000$ simulated cascades per point). Solid: final cascade direction $D_\infty$; dashed: majority $\sign(\Jbar_N)$. \textbf{Left}: a strong AI ($\alpha = 0.85 > q$) sets $D_\infty = c$ at every $k$ but wins the majority only for $k \leq \frac{N}{2}$ (shaded gap); a weak AI ($\alpha = 0.62 < q$) does neither. \textbf{Right}: the same two quantities against $\theta$, with classical no-AI baselines dotted.}
  \label{fig:relax-timing}
\end{figure}

If the indicator is attached only from position $k$, users $1, \dots, k-1$ run the classical cascade, so the AI prediction arrives against a public log-odds already locked at magnitude $\Lq$. A strong AI ($\La > \Lq$) overrides that lock: $D_\infty = c$ at every $k$ (Figure~\ref{fig:relax-timing}, solid). But the majority $\sign(\Jbar_N)$---the aggregate record an external observer reads---follows $c$ only when $c$ controls more than half the positions, $k \leq \frac{N}{2}$ (dashed). A late strong AI therefore redirects every user after $k$ yet leaves the recorded majority with the pre-AI herd, and a weak AI ($\La < \Lq$) can do neither.

\begin{figure}[t]
  \centering
  \includegraphics[width=0.62\linewidth]{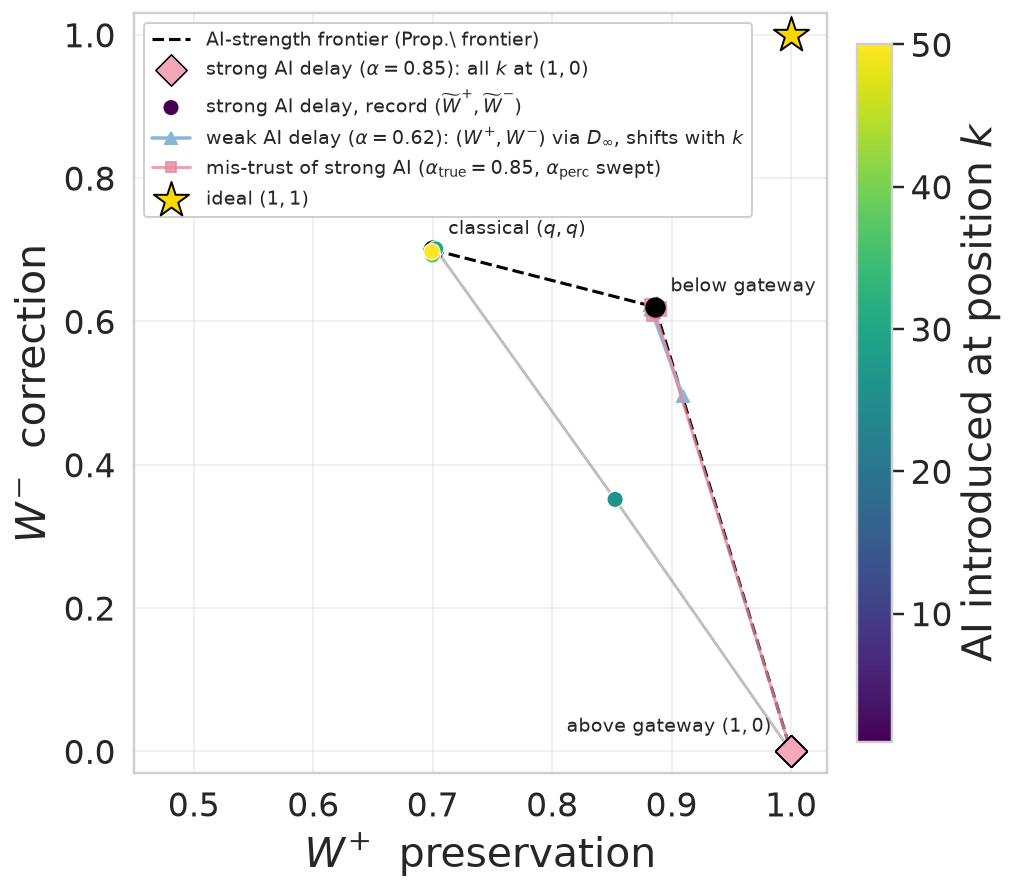}
  \caption{Both deployment levers on the welfare plane of Proposition~\ref{prop:frontier} ($q = 0.7$, $N = 50$): preservation $W^{+}$ against correction $W^{-}$, with the AI-strength frontier in black. Mis-trusting a strong AI ($\alpha_{\mathrm{true}} = 0.85$, $\alpha_{\mathrm{perc}}$ swept; pink) moves \emph{along} the frontier. Delaying the same AI to position $k$ leaves $(W^{+}, W^{-})$ at the above-Gateway $(1, 0)$ (diamond), while the record welfare $(\widetilde W^{+}, \widetilde W^{-})$, scored for $\sign(\Jbar_N)$ in place of $D_\infty$, slides inside the frontier toward the classical $(q, q)$ (colored by $k$), dominated by the below-Gateway point. Neither approaches the ideal $(1, 1)$.}
  \label{fig:relax-welfare}
\end{figure}

Neither lever escapes the frontier, but they fail differently, and for delayed exposure the failure depends on which object is scored. Miscalibrated trust moves the population \emph{along} the frontier: under-trusting a strong AI reproduces the below-Gateway point. Under delayed exposure the cascade direction itself never moves---a strong AI dictates $D_\infty$ from any position $k$, so $(W^{+}, W^{-})$ stays at the above-Gateway $(1, 0)$. What moves is the record. Because delay separates $D_\infty$ from $\sign(\Jbar_N)$, the record acquires welfare coordinates of its own,
\[
\widetilde W^{+} = \Pr\bigl(\sign(\Jbar_N) = \theta \mid c = \theta\bigr),
\qquad
\widetilde W^{-} = \Pr\bigl(\sign(\Jbar_N) = \theta \mid c \neq \theta\bigr),
\]
which coincide with $(W^{+}, W^{-})$ when the indicator is shown from the first position but not under delay. On this record welfare, delay slides $(\widetilde W^{+}, \widetilde W^{-})$ from $(1, 0)$ to the classical $(q, q)$ (Figure~\ref{fig:relax-welfare}), a path the below-Gateway point strictly dominates: judged by the aggregate a platform reports, a strong AI introduced late is worse on both coordinates than a weak AI shown from the start. Raising correction without surrendering preservation is therefore beyond either deployment lever; it requires the structural changes to the signal itself given in Section~\ref{sec:interventions}.

\section{Proofs}\label{app:proofs}

This appendix collects the proofs of all theorems, propositions, lemmas, and corollaries deferred from the main text. Proofs are listed in the order the corresponding results appear in the paper.

\subsection{Proof of Lemma~\ref{lem:cascade-structure} (Cascade structure)}\label{app:proof-regimes}
With $\ell_i = P_i + \lambda(s_i; q)$ and $\lambda(s_i;q) \in \{\pm\Lq\}$:
\emph{Case (a)}: $|P_i| < \Lq$ implies $\ell_i \neq 0$ for both signal realizations, so $J_i = 1 \iff s_i = 1$. Then $J_i$ perfectly reveals $s_i$, and the log-LR of $J_i$ given $\theta$ equals that of $s_i$, namely $\lambda(J_i; q)$.
\emph{Case (b)}: $P_i \geq \Lq$. If $s_i = 1$, $\ell_i = P_i + \Lq > 0$. If $s_i = 0$, $\ell_i = P_i - \Lq \geq 0$, and on the boundary $P_i = \Lq$ the tie rule (Appendix~\ref{app:tie-breaking}) yields $J_i = 1$. Hence $J_i = 1$ regardless of $s_i$, contributing zero log-LR.
\emph{Case (c)}: Symmetric.

\subsection{Proof of Theorem~\ref{thm:equilibrium} (Equilibrium structure)}\label{app:proof-equilibrium}
The initial condition $P_1 = \ell_0 + \lambda(c;\alpha)$ follows from~\eqref{eq:Pi-def} with $H_1 = \emptyset$. By Lemma~\ref{lem:cascade-structure}, in the learning region $\psi_i = \pm\Lq$ with sign equal to $\sign(2s_i - 1)$ via Lemma~\ref{lem:cascade-structure}(a), so the conditional mean of the increment is $\E[\psi_i \mid \theta] = (2q-1)\Lq \cdot \sign(2\theta - 1)$. In a cascade $\psi_i = 0$ by Lemma~\ref{lem:cascade-structure}, so $P_{i+1} = P_i$; hence the cascade regions $\{P \geq \Lq\}$ and $\{P \leq -\Lq\}$ are absorbing.

\subsection{Proof of Theorem~\ref{thm:cascade-genericity} (Immediate AI-Directed Absorption)}\label{app:proof-cascade-genericity}
By Corollary~\ref{cor:starting-point}, when $\alpha \geq q$ we have $P_1 = \lambda(c; \alpha)$. If $c = 1$, then $P_1 = \La \geq \Lq$, placing the chain in the up-cascade absorbing set; by Lemma~\ref{lem:cascade-structure}(b), user 1 ignores $s_1$ and chooses $J_1 = 1$. By Theorem~\ref{thm:equilibrium}, $P_2 = P_1 \geq \Lq$, so user 2 likewise chooses $J_2 = 1$. Induction gives $J_i = 1$ for all $i$. Symmetrically, $c = 0$ yields $J_i = 0$ for all $i$. Hence $J_i = c$ for all $i$, and no $s_i$ enters the public history.

\subsection{Proof of Proposition~\ref{prop:tau} (Cascade-Formation Time)}\label{app:proof-tau}
(a) is Theorem~\ref{thm:cascade-genericity}. (The measure-zero classical edge $\alpha = \tfrac{1}{2}$ is treated separately in Appendix~\ref{app:tie-breaking}.)

For (b), fix $\alpha \in (\frac{1}{2}, q)$, so that $0 < \La < \Lq$. By the symmetry $c \leftrightarrow 1-c$, $\theta \leftrightarrow 1-\theta$, $P \leftrightarrow -P$, it suffices to analyze the chain conditional on $c = 1$; the case $c = 0$ is identical with all signs reversed. The starting point is $P_1 = \La \in (0, \Lq)$, which lies in the learning region, so $\tau \geq 2$.

\emph{Step 1: the chain $(P_j)$ visits at most two non-absorbing states.} Define
\[
A \equiv \La, \qquad B \equiv \La - \Lq.
\]
Both lie in the learning region: $A \in (0, \Lq)$ and $B \in (-\Lq, 0)$, where the second inclusion uses $\La < \Lq$. From $A$ a learning-region step $\pm\Lq$ leads to either $\La + \Lq$ or $B$; the former satisfies $\La + \Lq \geq \Lq$, hence triggers an up-cascade ($U$). From $B$ a learning-region step leads to either $A$ or $\La - 2\Lq$; the latter satisfies $\La - 2\Lq \leq -\Lq$ \emph{precisely} because $\La < \Lq$, hence triggers a down-cascade ($D$). The transient state space conditional on $c = 1$ therefore reduces to $\{A, B\}$ with the four transitions
\[
A \to U,\; A \to B,\; B \to A,\; B \to D,
\]
and the topology of the chain is independent of $\alpha$ throughout $(\frac{1}{2}, q)$.

\emph{Step 2: transition probabilities.} In the learning region $\psi_i = \lambda(s_i; q)$, so $\Pr(\psi_i = +\Lq \mid \theta) = \Pr(s_i = 1 \mid \theta)$. Write
\[
p \equiv \Pr(s_i = 1 \mid \theta) = \begin{cases} q & \text{if } \theta = 1, \\ 1 - q & \text{if } \theta = 0. \end{cases}
\]
Conditional on $(c = 1, \theta)$, the four transitions have probabilities
\[
\Pr(A \to U) = p, \;\; \Pr(A \to B) = 1 - p, \;\; \Pr(B \to A) = p, \;\; \Pr(B \to D) = 1 - p,
\]
none of which depend on $\alpha$.

\emph{Step 3: $\Pr(\tau = 2)$.} Since $P_1 = A$ and $\tau = 2$ iff the first transition absorbs, $\Pr(\tau = 2 \mid c = 1, \theta) = \Pr(A \to U) = p$. Substituting $p = q$ for $c = \theta$ and $p = 1 - q$ for $c \neq \theta$ yields the displayed probabilities.

\emph{Step 4: $\E[\tau]$.} Let $\mu_A$ (resp.\ $\mu_B$) denote the expected number of transitions to absorption from $A$ (resp.\ $B$). First-step analysis gives
\[
\mu_A = 1 + (1 - p)\mu_B, \qquad \mu_B = 1 + p\,\mu_A.
\]
Substituting and solving,
\[
\mu_A \;=\; \frac{2 - p}{1 - p(1 - p)}.
\]
Since $\tau = 1 + (\text{transitions from } A)$,
\[
\E[\tau \mid c = 1, \theta] \;=\; 1 + \frac{2 - p}{1 - p(1 - p)}.
\]
Substituting $p = q$ and using $1 - q(1-q) = 1 - q + q^2$ gives $\E[\tau \mid c = \theta] = 1 + \frac{2 - q}{1 - q + q^2}$; substituting $p = 1 - q$ gives $\E[\tau \mid c \neq \theta] = 1 + \frac{1 + q}{1 - q + q^2}$. The difference is $\frac{2q - 1}{1 - q + q^2}$, strictly positive for $q > \frac{1}{2}$.

In none of these expressions does $\alpha$ appear: changing $\alpha$ within $(\frac{1}{2}, q)$ shifts the locations of $A$ and $B$ but preserves the four-arrow transition diagram, so the conditional law of $\tau$ given $(c, \theta)$ is exactly constant in $\alpha$.

\subsection{Proof of Theorem~\ref{thm:direction} (Cascade Direction under a Shared AI Signal)}\label{app:proof-direction}
Part (a) is immediate from Theorem~\ref{thm:cascade-genericity}. For (b), by the symmetry $c \leftrightarrow 1-c$, $\theta \leftrightarrow 1-\theta$, $P \leftrightarrow -P$, it suffices to condition on $c = 1$. Proposition~\ref{prop:tau} (Steps 1--2) reduces the transient state space to $\{A, B\} = \{\La,\, \La - \Lq\}$ with transitions $A \to U$ (up-cascade), $A \to B$, $B \to A$, $B \to D$ (down-cascade) having probabilities $p, 1-p, p, 1-p$ respectively, where $p = \Pr(s_i = 1 \mid \theta) \in \{q, 1-q\}$. Let $u_A$ and $u_B$ denote the probabilities of absorption at $U$ starting from $A$ and $B$. First-step analysis gives
\[
u_A = p + (1-p)\,u_B, \qquad u_B = p\, u_A,
\]
hence $u_A = \frac{p}{1 - p(1-p)} = \frac{p}{1 - p + p^{2}}$. Since $1 - p + p^{2}$ is invariant under $p \leftrightarrow 1-p$, substituting $p = q$ (case $c = \theta$) and $p = 1-q$ (case $c \neq \theta$) yields the displayed conditional probabilities. Neither expression depends on $\alpha$: changing $\alpha$ within $(\frac{1}{2}, q)$ shifts the locations of $A$ and $B$ but preserves the four-arrow transition diagram, which alone determines absorption probabilities.

\subsection{Proof of Corollary~\ref{cor:direction-marginal} (Marginal monotonicity in $\alpha$)}\label{app:proof-direction-marginal}
Average Theorem~\ref{thm:direction}(b) over $\theta \mid c$ using $\Pr(\theta = c \mid c) = \alpha$:
\[
\Pr(D_\infty = \sign(\lambda(c;\alpha)) \mid c) = \alpha \cdot \frac{q}{1 - q + q^{2}} + (1-\alpha) \cdot \frac{1-q}{1 - q + q^{2}} = \frac{\alpha q + (1-\alpha)(1-q)}{1 - q + q^{2}}.
\]
The numerator simplifies to $(1-q) + \alpha(2q-1)$, which is strictly increasing in $\alpha$ since $q > \frac{1}{2}$.

\subsection{Proof of Theorem~\ref{thm:spurious} (AI-Mediated Agreement and External Overconfidence)}\label{app:proof-spurious}

\noindent\emph{Information-aggregation bound.} We establish that, for every $N$,
\begin{equation}
I(\Jbar_N; \theta) \;\leq\; I\bigl((c, \sigma_{<\tau}); \theta\bigr) \;=\; I(c; \theta) + I(\sigma_{<\tau}; \theta \mid c).
\label{eq:info-bound}
\end{equation}
By the data-processing inequality, $I(\Jbar_N; \theta) \leq I((J_1, \ldots, J_N); \theta)$. For $j > \tau$, Theorem~\ref{thm:equilibrium} gives $\psi_j = 0$, so $J_j = J_\tau$ and the whole sequence is a deterministic function of $(J_1, \ldots, J_\tau)$; hence
\[
I((J_1, \ldots, J_N); \theta) = I((J_1, \ldots, J_\tau); \theta).
\]
By Lemma~\ref{lem:cascade-structure}, each $J_j$ with $j < \tau$ lies in the learning region and reveals $s_j$, while the forced judgment $J_\tau = \sign(P_\tau)$ is a function of $(\sigma_{<\tau}, c, \tau)$ and reveals nothing about $s_\tau$. Thus $(J_1, \ldots, J_\tau)$ is a function of $(\sigma_{<\tau}, c, \tau)$, and since $\tau$ is itself $\sigma((\sigma_{<\tau}, c))$-measurable, data-processing gives
\[
I((J_1, \ldots, J_\tau); \theta) \;\leq\; I((\sigma_{<\tau}, c); \theta).
\]
The chain rule together with conditional independence of $\{s_i\}$ from $c$ given $\theta$ yields
\[
I((\sigma_{<\tau}, c); \theta) = I(c; \theta) + I(\sigma_{<\tau}; \theta \mid c),
\]
establishing~\eqref{eq:info-bound}. (A conservative universal bound also holds, $I((c, \sigma_{<\tau}); \theta) \leq H(\theta) \leq \log 2$ for binary $\theta$, but the content of the theorem is that the right-hand side is fixed in $N$ through the finite-mean pre-cascade prefix, not merely capped at $\log 2$.)

For the prefix length, the worst case in Proposition~\ref{prop:tau}(c) is $\E[\tau \mid c \neq \theta] = 1 + \frac{1+q}{1-q+q^{2}}$ on $\alpha \in (\frac{1}{2}, q)$, which dominates the other regimes ($\E[\tau] = 1$ for $\alpha \geq q$ by part (a), $\E[\tau] = 2$ for $\alpha = \frac{1}{2}$ by part (b)). Hence
\[
\E[|\sigma_{<\tau}|] = \E[\tau - 1] \;\leq\; \frac{1+q}{1-q+q^{2}}
\]
uniformly in $\alpha$. For $\alpha \geq q$, $\tau = 1$ almost surely (Theorem~\ref{thm:cascade-genericity}), so the prefix $\sigma_{<\tau}$ is empty and $I(\sigma_{<\tau}; \theta \mid c) = 0$; since $J_1 = c$ deterministically, the judgment sequence carries no information about $\theta$ beyond $c$, and the bound holds with equality, $I(\Jbar_N; \theta) = I(c; \theta)$.

\medskip
\noindent\emph{External-overconfidence gap~\eqref{eq:bias}.}
The na\"ive observer assigns log-odds $\lambda(J_i; q) = (2J_i - 1) \cdot \Lq$ to each judgment, yielding $\ell^{\mathrm{naive}} = \ell_0 + \sum_i (2J_i - 1) \Lq$. Conditional on a cascade with consensus $1$ (the consensus-$0$ case is symmetric), the terms with $i \geq \tau$ all equal $+\Lq$ by Theorem~\ref{thm:equilibrium}, contributing $(N - \tau + 1) \cdot \Lq$; the first $\tau - 1$ terms contribute a quantity uniformly bounded by $(\tau - 1) \cdot \Lq$. Summing gives $N \cdot \Lq + O(\tau)$, and taking expectation projects to $N \cdot \Lq + O(1)$ on the consensus direction.

For the Bayes-correct observer, conditional on the realized judgment sequence she marginalizes over $(c, \sigma_{<\tau})$ using the model's likelihood. By the information-aggregation bound above, her posterior on $\theta$ has $I((c, \sigma_{<\tau}); \theta)$ uniformly bounded in $N$. In log-odds form,
\[
\bigl|\ell^{\mathrm{correct}}\bigr| \;\leq\; |\ell_0| + |\lambda(c;\alpha)| + (\tau - 1) \cdot \Lq \;\leq\; |\ell_0| + \La + (\tau - 1) \cdot \Lq
\]
pathwise, hence $\E[\,|\ell^{\mathrm{correct}}|\,] \leq |\ell_0| + \La + \E[\tau - 1] \cdot \Lq$. The signed projection $\E[\ell^{\mathrm{correct}} \cdot \sign(\mathrm{consensus})]$ is therefore a finite quantity, and subtracting the naive expression while absorbing the bounded Bayes-correct contribution into $O(1)$ yields~\eqref{eq:bias}.

The two benchmark constants follow by evaluating the Bayes-correct projection exactly under $\ell_0 = 0$. When $\alpha = \frac{1}{2}$, the first user reveals $s_1$ and every later judgment copies $J_1$, so the correct observer extracts exactly one private impression and $\ell^{\mathrm{correct}} \cdot \sign(\mathrm{consensus}) = \Lq$, giving the gap $N\Lq - \Lq$. When $\alpha \geq q$, $J_i = c$ for all $i$ reveals $c$ and nothing else, so $\ell^{\mathrm{correct}} = \lambda(c;\alpha)$ with projection $\La$, giving $N\Lq - \La$.

\subsection{Proof of Theorem~\ref{thm:cont-genericity} (Continuous-Indicator Immediate Absorption)}\label{app:proof-cont-genericity}
Identical to Theorem~\ref{thm:cascade-genericity} with $\Lambda(c)$ replacing $\lambda(c; \alpha)$.

\subsection{Proof of Proposition~\ref{prop:behavioral} (Behavioral threshold for first-user cascade)}\label{app:proof-behavioral}
Under the weighted update rule, user~1's posterior log-odds at decision time are $\ell_1^{\beta} = \ell_0 + \beta_{\mathrm{AI}} \cdot \lambda(c; \alpha) + \beta_{\mathrm{private}} \cdot \lambda(s_1; q)$, and the decision rule of Section~\ref{sec:model} (translated to the behavioral log-odds) is
\[
J_1 \;=\; \indicator\{\ell_1^{\beta} > 0\} + \indicator\{\ell_1^{\beta} = 0\} \cdot \indicator\{\ell_0 + \beta_{\mathrm{AI}} \cdot \lambda(c;\alpha) \geq 0\},
\]
breaking ties in the direction of the public (AI-indicator-plus-prior) log-odds, which generalizes the $\sign(P_i)$ tie rule of Section~\ref{sec:model}. With $\ell_0 = 0$, the cascade-genericity argument of Theorem~\ref{thm:cascade-genericity} carries through with the threshold comparison $|\beta_{\mathrm{AI}} \cdot \lambda(c;\alpha)|$ vs.\ $\beta_{\mathrm{private}} \cdot \Lq$ in place of $|\lambda(c;\alpha)|$ vs.\ $\Lq$. When this comparison favors the AI-indicator side, user 1's judgment is determined by $c$ regardless of $s_1$: if $c = 1$, $\beta_{\mathrm{AI}} \La \geq \beta_{\mathrm{private}} \Lq$ gives $\ell_1^\beta \geq 0$ for both signal realizations, and the tie rule selects $J_1 = 1$ at equality; if $c = 0$, the symmetric argument selects $J_1 = 0$. Theorem~\ref{thm:equilibrium}'s absorbing property of the cascade region (which depends only on the increment structure of subsequent $P_j$, not on the magnitude of $P_1$) propagates the cascade.

\end{document}

%% file: sections/01-introduction.tex
\section{Introduction}
Misinformation on social platforms remains a critical challenge. It spreads
through user interactions: users judge the veracity of news stories and express
those judgments through comments, ratings, flags, or sharing decisions
~\cite{pennycook2021psychology}. To combat misinformation, platforms increasingly
attach AI-based credibility indicators to news stories, such as predictions about
whether a story contains false information
~\cite{jia2022understanding,seo2019trust,jiang2026all}. This creates a familiar human--AI decision-making problem: users' reliance on AI predictions determines whether they can correctly judge the veracity of online content and engage with accurate information.

However, AI-based credibility indicators work differently from standard one-shot
AI advice. A user who sees the AI prediction may also see how earlier users
reacted after seeing the same AI signal, and her own judgment may then become
part of the public record seen by later users
~\cite{bikhchandani1992theory,banerjee1992herd}. Existing human--AI
decision-making theory has largely focused on whether an individual user revise her initial judgment after receiving AI advice, leaving under-explored how AI advice reshapes a sequence of user decisions, such as news veracity judgments in the presence of AI. In this setting, individual reliance on AI can change what later users learn from the public history.

Fortunately, formally characterizing sequential human behavior is not a new problem: social learning theories have studied how people update and act on information from others for decades. We therefore introduce a new lens for studying sequential human engagement with AI-based credibility indicators by extending this existing framework: Bayesian belief updating with a shared AI signal. In this view, each user updates her belief about a story by combining three signals: her private impression of the story, the AI credibility signal, and the judgments of earlier users~\cite{tenenbaum2011grow,griffiths2006optimal,steyvers2022bayesian}. Representing these signals on a common log-odds scale allows us to formally describe how AI predictions affect sequential veracity judgments.

What core mechanism does a shared AI signal introduce into a sequence of
judgments and the resulting public history? Classical social-learning models
show that earlier judgments accumulate into public belief until a cascade forms,
after which later actions no longer reveal private information
~\cite{bikhchandani1992theory,banerjee1992herd}. The AI signal enters this
process differently. Because every user sees the same signal, it adds a fixed
amount to each user's belief and sets the starting point of the sequence. This difference gives a simple condition, which we call the \emph{Gateway}. The
Gateway compares the evidence from the AI prediction with the evidence from one
user's private impression. In accuracy terms, it compares the AI accuracy
\(\alpha\) with the human private-impression accuracy \(q\). In log-odds terms,
one AI indication is worth \(k^{*}=\La/\Lq\) private impressions, so the Gateway
is \(k^{*}\geq1\). When the AI is at least as strong as a user's private
impression \((\alpha\geq q)\), the first user follows the AI regardless of her
own impression. Her judgment reveals no new private information to the next user,
so the same logic repeats. The crowd is absorbed into the AI prediction, right or
wrong, and private impressions never enter the public record. When the AI is weaker \((\alpha<q)\), early users can still act on their own
private impressions before a cascade forms. These impressions can correct a wrong
AI prediction, but they can also overturn a correct one. Conditional on the AI's
realized indication for a given story, the subsequent dynamics depend on \(q\),
not on the AI's overall accuracy \(\alpha\). Higher AI accuracy only changes how
often the process begins with a correct indication. Once that indication is
fixed, it does not make the crowd more effective at pooling its own information
or more strongly led toward the truth.

The Gateway mechanism has both benefits and risks. Externally, it changes what population statistics mean. In the presence of a shared AI signal, crowd agreement becomes more dependent than it appears: a Bayes-correct observer counts only the shared AI signal and the finite pre-cascade private impressions, while a naive observer may mistake repeated agreement for many independent human judgments. Internally, the same mechanism changes the welfare of following the AI. When the AI is correct, absorption can preserve the correct prediction by filtering out noisy private impressions; when the AI is wrong, the same absorption blocks the corrective impressions that could overturn it. Improving AI accuracy moves the system toward the better side of this trade-off, but it does not eliminate the frontier: stronger AI also makes wrong AI predictions harder for the crowd to correct.


We calibrate the model against data from a prior human-subject experiment in which users judge news veracity in the presence of an AI indicator that outperforms human users. We first benchmark users as fully calibrated Bayesian agents and find though users are significantly affected by AI, there are deviations from immediate cascades. We then relax the decision rule, estimating behavioral weights on each signal, and find that the average user sits in the learning regime of the Gateway, treating the AI prediction equivalent to $3$---$6$ peer judgments. Finally, fitting a mixture model over subjects reveals two distinct behavioral components: a majority who under-weight the AI and never cascade on it alone, and a cascade-prone minority who over-weight it and cascade immediately. We also use simulations to examine two implications beyond the calibration data.
First, by sweeping over combinations of the AI's true accuracy and users'
perceived AI accuracy, we show that over-reliance a weak AI is more damaging than
under-reliance on a strong one. Second, simulations of structural interventions show
that diversification and expert review improve the preservation-correction
trade-off in different ways. Diversifying the AI signal across users is more
reliable across AI strengths and behavioral weightings, while triggering
expert review in slowly forming cascades is useful mainly when users remain in the learning regime.

To summarize, we make three contributions. First, we introduce a social-learning
lens for information spread with AI-based credibility indicators, modeling the
indicator as a platform-deployed shared signal that links individual reliance to
crowd-level dependence. Second, we derive interpretable consequences of this
shared-signal view: a Gateway condition for when independent human evidence stops
entering the public record, an information ceiling for AI-shaped consensus, and a
preservation--correction trade-off.
Third, we calibrate a behavioral version of the model on human-subject sequential judgment data, showing average users weight AI below their own impressions yet above several peer
judgments, while the AI treatment reveals a cascade-prone behavioral component that crosses the Gateway. We hope this work provides a starting point for extending
human-AI interaction theory to crowd-AI interaction.

%% file: sections/02-related-work.tex
\section{Related Work}\label{sec:related}


\subsection{Online information spread: sequential veracity judgments on misinformation}

Misinformation has been a critical issue~\cite{jerit2020political,aimeur2023fake}. It spreads through platforms as a sequence of human judgments: users encounter a story, assess its veracity, and react---and false content travels farther and faster than true content through exactly this relay~\cite{vosoughi2018spread, pennycook2021shifting, shao2016hoaxy, lu2022effects, liu2026autoredtrader, lu2026large}. Two empirical facts shape the environment. First, individual truth discernment is limited; people separate true from false headlines only modestly better than chance \citep{pennycook2021psychology}. Second, users do not judge in isolation: they see the reactions of those before them---votes, shares, comments---and such social signals causally shift later judgments, as shown by field experiments on rating herding and social influence bias~\cite{shao2016hoaxy, muchnik2013social, lu2022effects, jahanbakhsh2020experimental}. Platforms now fold these judgments into crowd-facing statistics, from crowdsourced credibility ratings to community fact-checking notes~\cite{borenstein2025can, prollochs2022community, bhuiyan2020investigating, chuai2024did}, so each user's veracity judgment is simultaneously her own decision and social evidence for everyone after her. Our model takes this sequential structure as primitive and perturbs it with one design change: an AI signal at every position.

\subsection{Human-AI decision making in information processing}

A large body of work in human-AI interaction studies how AI assistance changes individual decisions: when people accept algorithmic advice, how they calibrate reliance on it, and how interface and explanation design shift that reliance~\cite{bansal2019beyond, bansal2021does, lai2023towards, lai2019human, guo2024decision, lu2021human}. People discount AI predictions generally \citep{yaniv2004advice, bailey2023meta}, but show over-reliance (algorithm appreciation, \citealt{logg2019algorithm, hou2021expert}) and sometimes under-reliance due to different cognitive factors (algorithm aversion, \citealt{dietvorst2015algorithm, dietvorst2018overcoming}). For misinformation specifically, AI-based credibility indicators and warning labels measurably move individual belief and sharing decisions~\cite{lu2022effects, epstein2022explanations,jahanbakhsh2023exploring,dietvorst2018overcoming,lu2025understanding}. This line of work provides empirical and theoretical grounding to individual human-AI decision-making where human engagement is determined primarily by the decision-maker and the AI~\cite{yin2019understanding,guo2024decision,li2024decoding}. In information spread, however, reliance on AI does not remain an individual
decision. It can enter the public history through visible user judgments and
shape what later users learn from the crowd. This sequential dimension of
human-AI decision-making remains under-explored, especially from a analytical
perspective.

\subsection{Belief update: a theoretical lens to uniformly model both}

Analyzing social influence and AI influence in one system requires a common currency, and Bayesian belief update in log-odds form supplies it: every signal---a private impression, an AI prediction, an earlier user's judgment---enters as an additive evidence contribution. This is the machinery of observational learning in computational economics: information cascades \citep{bikhchandani1992theory, banerjee1992herd, glenski2017rating, prollochs2023mechanisms}, their limits under unbounded private beliefs \citep{smith2000pathological,horvat2023hidden,almaatouq2020social}, observation networks \citep{acemoglu2011bayesian}, price aggregation in markets \citep{vives1993fast,brahma2012bayesian}, and word-of-mouth learning with a public signal, the closest precedent to ours \citep{banerjee2004word,cherng2022understanding}. Behavioral departures from the Bayesian rule---conservatism~\cite{biswas2026belief,karduni2020bayesian,markant2023data}, noisy quantal choice~\cite{wright2019level,milec2021complexity,feng2024rationality}---extend the same currency to measured human behavior, which is exactly the bridge our behavioral calibration crosses.

Prior social-learning models supply the machinery for common signals; we put it to a new use---reading AI credibility indicators as platform-deployed signals whose effects propagate through sequential human judgments and reshape what crowd statistics mean. This reframing, rather than a new equilibrium concept, is where the core mechanism and analysis in Sections~\ref{sec:main} and~\ref{sec:empirical} come from.

%% file: sections/03-model.tex
\section{Model and Dynamics}\label{sec:model-dynamics}

We study a typical setting of information spread in the presence of an
AI-based credibility indicator. A user, Alice, is browsing news on a social
media platform. Seeing a news story, she first forms her own impression of
whether it is true; alongside the story, she also observes two public cues: an
AI-based credibility indicator attached to the content, and the judgments left
by earlier users (e.g., potentially through comments, upvotes, and downvotes). She then makes her own veracity judgment and leaves it in the comment section, where it becomes
part of the record shown to later users. Our research question is how to model
this entire information-spread process, where a sequence of such users, each
judging under the same AI indicator, through a single theoretical lens.

\subsection{Cascade Model with AI-based Credibility Indicator}\label{sec:model}

We deliberately use a minimal extension of the classical BHW information
cascade~\cite{bikhchandani1992theory, banerjee1992herd}. The only new element is a
common AI credibility signal: one prediction $c$, computed once for the story and
shown unchanged to every user.\footnote{We use \emph{user} and \emph{judgment}
throughout for what the cascade literature calls the \emph{agent} and her
\emph{action}.} The minimality is the point: any change in the public judgment
record can then be traced to the shared-signal role of the AI indicator.
A news story has unknown binary veracity
$\theta \in \{0 = \textsc{Fake},\, 1 = \textsc{Real}\}$, with common prior
$\pi_0 = \Pr(\theta = 1)$, and $N$ users, indexed $i = 1, \ldots, N$, judge it
one after another. Each user $i$ receives three signals:
\begin{itemize}
\item her \textbf{private impression} $s_i \in \{0,1\}$, her independent read
of the news veracity, with accuracy $\Pr(s_i = \theta \mid \theta) = q >
\tfrac{1}{2}$, drawn independently across users given $\theta$;
\item the \textbf{common AI signal} $c \in \{0,1\}$ produced by the AI
credibility indicator, with accuracy $\Pr(c = \theta \mid \theta) = \alpha >
\tfrac{1}{2}$, computed once per story and shown unchanged to every user;
\item the \textbf{public history} $H_i = (J_1, \ldots, J_{i-1})$, with $H_1 =
\emptyset$: the judgments earlier users left alongside the news, e.g., through
comments.
\end{itemize}

As in the classic cascade setting, private impressions stay hidden, and only
judgments are observable to later users. User $i$ chooses a judgment $J_i$
that minimizes $0$--$1$ loss against the story's true veracity, so her
interaction with the three signals is a Bayesian belief update. Taking $s_i$
and $c$ to be mutually independent given $\theta$, her posterior log-odds
$\ell_i$ are additive in the contributions of the signals she
sees~\cite{griffiths2006optimal, steyvers2022bayesian}:
\begin{equation}
\ell_i \;=\; P_i + \lambda(s_i;q),
\qquad
P_i \equiv \ell_0 \,+\, \lambda(c;\alpha) \,+\, \sum\limits_{j<i} \psi_j(J_j),
\label{eq:Pi-def}
\end{equation}
and $J_i = \indicator\{\ell_i > 0\}$. Here, for a binary signal of accuracy
$\rho \in (\tfrac{1}{2}, 1)$, $\lambda(x;\rho) = \pm L_\rho$ contributes to the
log-odds of $\theta = 1$, with the sign set by the value of $x$ and the
magnitude set by $L_\rho \equiv \log\frac{\rho}{1-\rho} > 0$; the private
impression and the AI prediction thus contribute $\pm\Lq$ and $\pm\La$,
respectively. Here $\psi_j(J_j)$ is the log-likelihood ratio that later users infer from
observing judgment $J_j$, given the public log-odds $P_j$ under which user $j$
acted:

\[
\psi_j(J_j; P_j)
=
\log
\frac{\Pr(J_j \mid \theta = 1, P_j)}
     {\Pr(J_j \mid \theta = 0, P_j)} .
\]
The posterior log-odds thus decompose into the \textbf{public log-odds} $P_i$,
common to every user at position $i$, and the private contribution
$\lambda(s_i;q)$. How this scalar evolves is the subject of
Section~\ref{sec:dynamics}.

We impose several technical conventions. First, ties at $\ell_i=0$---a measure-zero boundary event---are broken toward the public component, so at the boundaries $P_i=\pm\Lq$ the user chooses $J_i=\indicator\{P_i\geq 0\}$. The supplementary material analyzes both this conventional rule and a random tie-break and shows the choice bites only at the measure-zero classical edge $\alpha=\tfrac{1}{2}$, leaving every main conclusion unchanged. Second, we assume $q,\alpha\in(\tfrac{1}{2},1)$ throughout the main analysis, so both $\Lq$ and $\La$ are finite; the boundary value $\alpha=\tfrac{1}{2}$ (the classical no-AI limit) is itself a measure-zero edge, and we relegate it, together with the other measure-zero events, to the supplementary material. The limiting case $\alpha=1$, in which the AI model predicts $\theta$ perfectly, is degenerate and is therefore excluded from the main analysis. Finally, we focus on the symmetric prior $\pi_0 = \tfrac{1}{2}$ (so $\ell_0 = 0$) and treat the asymmetric case separately in the supplementary material, thus simplifying the public log-odds to $P_i \equiv \lambda(c;\alpha) \,+\, \sum\limits_{j<i} \psi_j(J_j)$. All boundary statements involving weak inequalities below (e.g., $\alpha\geq q$) adopt this tie-breaking convention; under random tie-breaking the weak inequalities become strict ($\alpha>q$), and the boundary cases $\alpha=q$ and $\alpha=\tfrac{1}{2}$ are analyzed in full in the supplementary material.



%% file: sections/04-dynamics.tex
\subsection{Public Log-Odds Dynamics: Public History and the AI Signal}\label{sec:dynamics}

We now characterize how the public log-odds evolve as users sequentially add
their judgments to the public history. At the moment user $i$ acts, the public
component of her posterior log-odds is $P_i = \ell_0 + \lambda(c;\alpha) +
\sum_{j<i}\psi_j(J_j)$; once her judgment joins the public history, the next
user faces
\[
P_{i+1} \;=\; P_i + \psi_i(J_i).
\]
In this recursion, the only newly added term is the contribution of user $i$'s
judgment. The AI contribution $\lambda(c;\alpha)$, by contrast, generates no
new term: because the same prediction $c$ is shown to every user for the same
story, it enters the public log-odds once and is carried forward throughout
the sequence.

This is the same public-belief reduction used in the classical cascade model:
the common AI signal changes the level of the public log-odds but not
its Markovian structure, since conditional on $\theta$ and $c$ the next
public log-odds depends on the past only through $P_i$. We can
therefore characterize the transition rule in the classical way, by asking
when user $i$'s judgment reveals her private impression: since $\ell_i = P_i +
\lambda(s_i;q)$, the judgment is informative exactly when the public component
is not strong enough to dominate one private impression, i.e., when $|P_i|$
has not reached the boundary $\Lq$.

\begin{lemma}[Cascade structure]\label{lem:cascade-structure}
For each user $i$:
\begin{enumerate}[label=(\alph*)]
\item \textbf{Learning regime.} If $|P_i| < \Lq$, then $J_i = 1
\iff s_i = 1$, and $\psi_i(J_i) = \lambda(J_i;q)$.
\item \textbf{Up-cascade.} If $P_i \geq \Lq$, then $J_i = 1$ regardless of
$s_i$, and $\psi_i(J_i) = 0$.
\item \textbf{Down-cascade.} If $P_i \leq -\Lq$, then $J_i = 0$ regardless of
$s_i$, and $\psi_i(J_i) = 0$.
\end{enumerate}
\end{lemma}

Combining this dichotomy with the Markovian representation yields the full
equilibrium of the public log-odds process.

\begin{theorem}[Equilibrium structure]\label{thm:equilibrium}
Conditional on $\theta$ and $c$, the public log-odds process $(P_i)$ is a
Markov chain with initial value
\[
P_1 = \ell_0 + \lambda(c;\alpha).
\]
It has a transient learning region $(-\Lq, \Lq)$ and two absorbing cascade
regions, $\{P \geq \Lq\}$ and $\{P \leq -\Lq\}$. While $P_i \in (-\Lq, \Lq)$,
the process moves by $+\Lq$ after judgment $J_i = 1$ and by $-\Lq$ after
judgment $J_i = 0$; equivalently,
\[
\E[P_{i+1} - P_i \mid \theta, P_i] = (2\theta - 1)(2q - 1)\Lq .
\]
Once $P_i$ reaches either cascade region, the process is absorbed.
\end{theorem}

Being common across users, the AI contributes no transition term; its only role is the initial condition $P_1 = \lambda(c;\alpha)$, so whether the sequence begins in the learning region or with an immediate cascade comes down to whether $\La \geq \Lq$.

With the guaranteed absorption by Theorem~\ref{thm:equilibrium}, two
statistics summarize the procedure. The first is
the \textbf{cascade-formation time}
\[
\tau = \inf\{\, i \geq 1 : |P_i| \geq \Lq \,\},
\]
the first position at which the public log-odds are absorbed. The second is the
\textbf{final cascade direction}
\[
D_\infty = \lim_{i \to \infty} \operatorname{sign}(P_i) \in \{-1, +1\},
\]
well-defined by eventual absorption. The comparison between $\La$ and $\Lq$, which
Section~\ref{sec:main} names the Gateway, already settles the
strong-AI case: $\tau = 1$ and $D_\infty =
\operatorname{sign}(\lambda(c;\alpha))$ when $\alpha \geq q$. Below it, both
statistics are random, and characterizing them is the work of the next
section.

%% file: sections/05-main-results.tex
\section{Main Results}\label{sec:main}

As we illustrated, one comparison translates individual reliance on the AI into a crowd-level condition---whether later judgments can still reveal independent human evidence. We call this comparison the \emph{Gateway}: the AI
contribution \(\La\) measured against one private-impression contribution
\(\Lq\), equivalently the AI accuracy \(\alpha\) measured against the human
accuracy \(q\).

\begin{corollary}[Gateway]\label{cor:starting-point}
With symmetric prior \(\ell_0=0\), the initial public log-odds \(P_1\) lies in
the learning region if and only if \(\alpha<q\). If \(\alpha\geq q\), \(P_1\)
lies in a cascade region, with the direction determined by the AI prediction
\(c\) (the equality case \(\alpha=q\) relies on the tie-breaking convention of
Section~\ref{sec:model}).
\end{corollary}

\begin{remark}[Evidence weight of one AI indication]\label{rmk:equivalent-population}
In log-odds, one AI indication is worth \(k^{*}=\La/\Lq\) independent private
impressions, so the Gateway \(\alpha\geq q\) is exactly \(k^{*}\geq 1\)---the AI
worth at least one private impression.
\end{remark}

\begin{figure}[htbp]
  \centering
  \includegraphics[width=\linewidth]{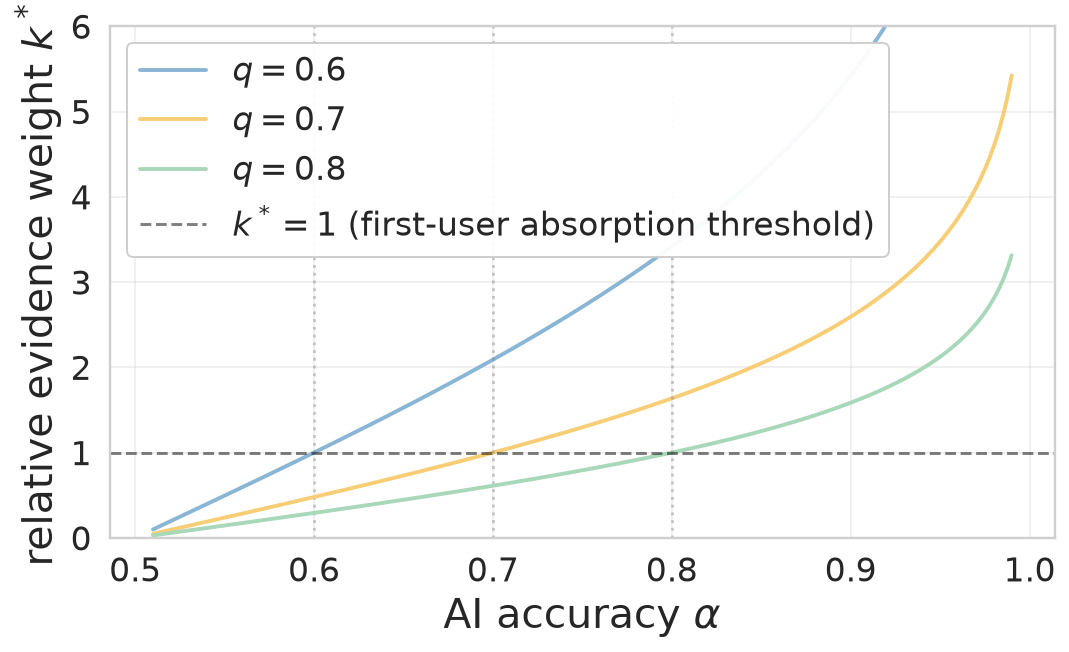}
  \caption{Evidence weight of the shared AI signal relative to one private impression (Remark~\ref{rmk:equivalent-population}): the relative log-odds weight $k^*(\alpha) = \frac{L_\alpha}{L_q}$ of the shared AI credibility signal versus one private impression, for $q \in \{0.6, 0.7, 0.8\}$. The boundary $k^*=1$ marks the first-user absorption threshold: at or above it, the AI signal can dominate one private impression, with each curve crossing $k^*=1$ exactly at $\alpha = q$.}
  \label{fig:equivalent-population}
\end{figure}

Once a cascade forms, later judgments reveal no additional private impressions, so the
public history contains only bounded information about \(\theta\), even as the
visible sequence of agreeing judgments can create excessive apparent support for
the cascade direction. Viewed as a shared signal, the indicator has three implications for reading a crowd's record: when independent human evidence stops entering it, how much it can ever contain, and when preserving an AI-correct cascade trades off against correcting an AI error. ~\footnote{Unless stated otherwise, all results condition on the observed AI
prediction \(c\). All tie-breaking cases and measure-zero boundary cases are
handled in the supplementary material. The conclusions are unchanged under either
conventional one-sided tie-breaking or random tie-breaking.}

\subsection{Above the Gateway: immediate AI-directed cascades}

We first consider the strong-AI side of the Gateway. When \(\alpha\geq q\), the
initial public log-odds is already at or beyond the relevant cascade boundary,
with the sign selected by the AI prediction.

\begin{theorem}[Immediate AI-Directed Absorption]
\label{thm:cascade-genericity}
If \(\alpha\geq q\), then for either AI prediction \(c\), each user $i$ makes the same judgment as the AI prediction, regardless of whether it is correct:
\[
J_i=c \qquad \text{for all } i\geq 1.
\]
That is, the cascade forms at the first user, \(\tau=1\), and the final
cascade direction will be the same as the AI prediction:
\[
D_\infty=\sign(\lambda(c;\alpha)).
\]
Thus, no user's private impression is revealed through the public history.\footnote{The boundary case \(\alpha=q\) uses the tie-breaking convention of Section~\ref{sec:model}; under random tie-breaking the statement holds for \(\alpha>q\).}
\end{theorem}

Correctness of the crowd thus reduces to correctness of the AI: if \(c=\theta\), the public history locks onto the truth; if \(c\neq\theta\), it locks onto the wrong direction. The proof is in the supplementary material.\footnote{Meanwhile, if the prior is asymmetric $(i.e., \ell_0 \neq 0)$, then the first-user cascade threshold depends on both the prior and the AI prediction ($P_1=\ell_0+\lambda(c;\alpha) $), with a Gateway threshold
\(\alpha\geq\alpha^*(q,\ell_0,c)\). The full
statement and figure are deferred to the supplementary material.}

\subsection{Below the Gateway: biased but not determined cascades}

We now consider the below-Gateway regime, \(\frac12<\alpha<q\). Here the AI
prediction is informative, but it is not strong enough to place \(P_1\) in a
cascade region. It shifts the initial public log-odds toward the direction of the AI prediction, while early private impressions can still enter the public history. Thus the AI prediction \textit{biases} both (1) \textit{when} the cascade forms and (2) \textit{which} direction it eventually takes, but it does not determine either one by itself.

\noindent\textbf{Cascade formation time} Particularly, the below-Gateway regime has a simple two-state structure before absorption. Changing
\(\alpha\) moves the transient public states, but as long as \(\alpha\in(\frac12,q)\), it does not change which private impressions move the
process between those states or into a cascade region. Those moves are governed
by \(q\). This gives the following timing result.

\begin{proposition}[Cascade-Formation Time]
\label{prop:tau}
Let $\tau$ be the cascade-formation time:
\[
\tau=\inf\{j\geq 1: |P_j|\geq \Lq\}
\] 
\begin{itemize}
\item[(a)] If \(\alpha\geq q\), then \(\tau=1\), as in
Theorem~\ref{thm:cascade-genericity}.

\item[(b)] If \(\alpha\in(\frac12,q)\), then for fixed \(c\) and \(\theta\), the value of $\tau$ depends on \(q\) and
on whether \(c=\theta\), but not on the value of \(\alpha\) within the
interval. A correct AI prediction yields a shorter expected cascade-formation time
than an incorrect one, with the difference governed by \(q\). Closed forms for
\(\Pr(\tau=2)\) and \(\mathbb{E}[\tau]\) are in
the supplementary material.
\end{itemize}
\end{proposition}

\noindent \textbf{Cascade Direction} The same below-Gateway structure also determines the final cascade direction. Let \(D_\infty\) denote the final cascade direction: \(D_\infty=+1\) if the
process is absorbed in an up-cascade, and \(D_\infty=-1\) if it is absorbed in a
down-cascade. Equivalently, by Theorem~\ref{thm:equilibrium},
\[
D_\infty=\lim_{j\to\infty}\sign(P_j).
\]
Particularly, below the Gateway, early private impressions can still correct a wrong AI prediction or overturn a correct one.
\begin{theorem}[Cascade Direction under a Shared AI Signal]
\label{thm:direction}
\mbox{}
\begin{itemize}
\item[(a)] If \(\alpha\geq q\), then for any fixed AI prediction \(c\),
\[
D_\infty=\sign(\lambda(c;\alpha)).
\]

\item[(b)] If \(\frac12<\alpha<q\), then, for fixed \(c\) and \(\theta\),
\[
\Pr\!\left(D_\infty=\sign(\lambda(c;\alpha))\mid c,\theta\right)
=
\begin{cases}
\dfrac{q}{1-q+q^2}, & c=\theta,\\[6pt]
\dfrac{1-q}{1-q+q^2}, & c\neq\theta.
\end{cases}
\]
\end{itemize}
\end{theorem}

This independence from \(\alpha\) holds only below the Gateway. Within \((\frac12,q)\), \(\alpha\) changes how often \(c=\theta\), but conditional on
\(c\) and \(\theta\), pre-cascade learning is driven by private impressions with
accuracy \(q\). At \(\alpha=q\), this learning phase vanishes: the initial
public log-odds reaches the cascade boundary, and the AI selects the final
direction from the first user onward.

Only judgments before cascade formation can reveal private impressions. The next
section uses this finite pre-cascade prefix to show why a large AI-shaped
consensus need not contain much independent human information about \(\theta\). The Gateway earns its place as a mechanism rather than a destination: it marks where the public record stops adding independent human evidence, which is what lets AI-shaped agreement mislead an outside observer.

\begin{theorem}[AI-Mediated Agreement and External Overconfidence]
\label{thm:spurious}
The Bayes-correct observer's posterior log-odds are bounded in expectation
uniformly in \(N\),
\[
\E\!\left[\left|\ell_{\mathrm{ext}}^{\mathrm{correct}}\right|\right]
\leq
|\ell_0|+\La+\E[\tau-1]\,\Lq,
\]
and the naive observer's excess log-odds in the final consensus direction grow
linearly,
\begin{equation}
\E\Bigl[
\bigl(\ell_{\mathrm{ext}}^{\mathrm{naive}}
-
\ell_{\mathrm{ext}}^{\mathrm{correct}}\bigr)D_\infty
\Bigr]
=
N\Lq+O(1),
\label{eq:bias}
\end{equation}
where the \(O(1)\) term is uniformly bounded in \(N\). Taking \(\ell_0=0\), the
constant is pinned down by the Gateway:
\begin{itemize}
\item[(a)] \textbf{Above the Gateway} (\(\alpha\geq q\), so \(\tau=1\) and
\(J_i=c\) for all \(i\)):
\[
\E\Bigl[
\bigl(\ell_{\mathrm{ext}}^{\mathrm{naive}}
-\ell_{\mathrm{ext}}^{\mathrm{correct}}\bigr)D_\infty
\Bigr]
=
N\Lq-\La.
\]
The only independent evidence the Bayes-correct observer can credit is the
shared AI prediction \(c\); no private impression is revealed.

\item[(b)] \textbf{Below the Gateway} (\(\tfrac12<\alpha<q\), so \(\tau\geq2\)):
the \(O(1)\) term equals the bounded evidence in \(c\) together with the finite
pre-cascade prefix \(\sigma_{<\tau}\), and is uniformly bounded in \(N\) by
\(\La+\E[\tau-1]\,\Lq\).
\end{itemize}
\end{theorem}

The two cases show that AI changes the source instead of the rate of external overconfidence. The naive observer counts every repeated judgment as a
fresh private signal, so excess evidence grows at slope \(\Lq\) in all cascade regimes. What differs is the bounded evidence being over-counted: below the
Gateway, it is \(c\) plus the finite pre-cascade prefix; above the Gateway, it is only \(c\), because no private impression is revealed. Particularly, this is a risk introduced by the presence of AI. A platform may display many users judging a news story (e.g., up-voting). If those users first saw
the same AI prediction, the statistic can be read as independent crowd validation, even though it reflects the impact of AI. The consensus is real, but the independence is further limited or even does not exist at all.

\begin{figure}[htbp]
  \centering
  \includegraphics[width=\linewidth]{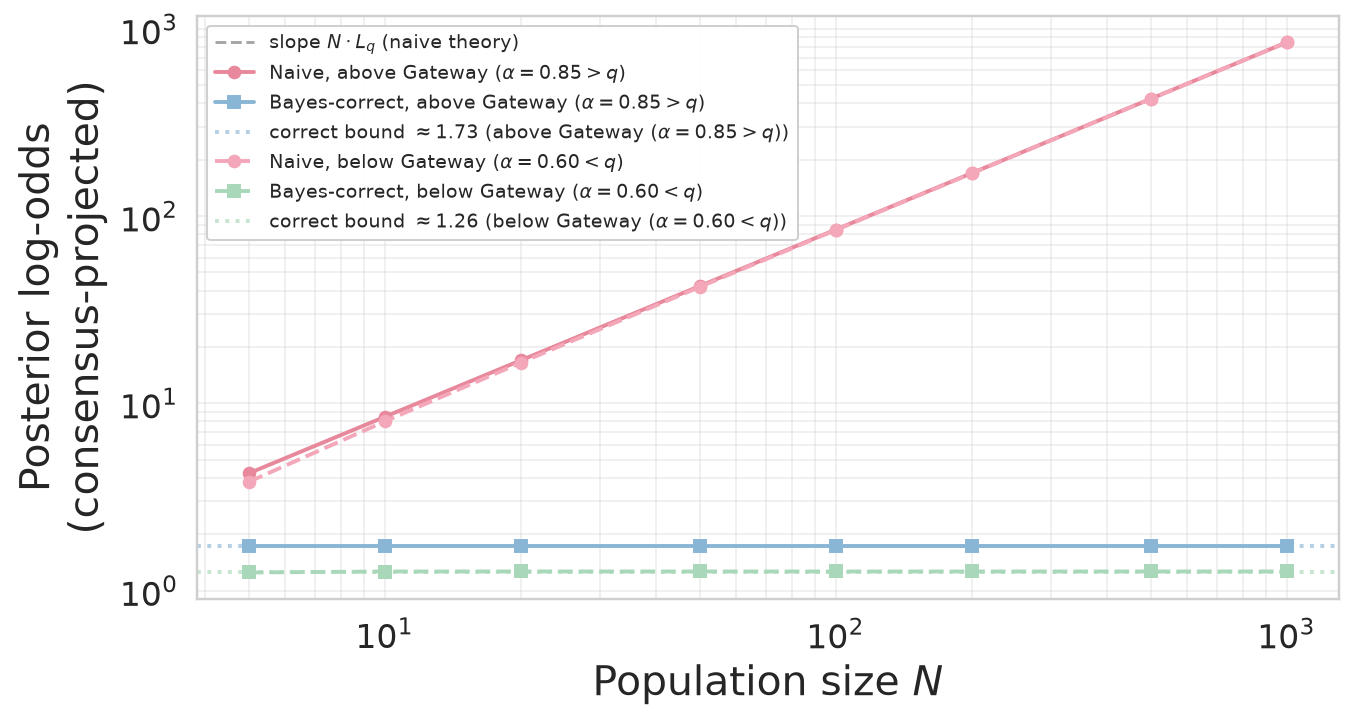}
  \caption{Spurious external confidence (Theorem~\ref{thm:spurious}), $q = 0.7$. Expected posterior log-odds projected onto the consensus direction, plotted against $N$ on a log--log scale. For both $\alpha = 0.85 > q$ (above Gateway, solid) and $\alpha = 0.60 < q$ (below Gateway, dashed), the na\"ive observer's curve grows as $N \cdot \Lq$ without bound; the Bayes-correct observer's curve stays bounded---at $L_\alpha \approx 1.73$ above the Gateway and at the cascade-formation threshold ${\approx}1.26$ below it.}
  \label{fig:spurious-confidence}
\end{figure}

\subsection{AI-based credibility indicator to combat misinformation: protecting correct cascades, correcting wrong ones}
\label{sec:welfare}

With an AI-based credibility indicator, the cost of absorption is ambiguous. Once the process is absorbed, later private impressions no longer enter the
public history, but this can help or harm depending on whether the AI is correct. If \(c=\theta\), absorption filters out noisy impressions that could pull users
away from the truth; if \(c\neq\theta\), it blocks the corrective impressions that could overturn the AI's mistake. We therefore evaluate the indicator along
two coordinates, conditional on AI correctness:
\[
W^{+}=\Pr(D_\infty=\theta\mid c=\theta),
\qquad
W^{-}=\Pr(D_\infty=\theta\mid c\neq\theta),
\]
the probability that the cascade direction is correct when the AI is right
(\(W^{+}\), \emph{preservation}) and when it is wrong (\(W^{-}\),
\emph{correction}), with ideal point \((W^{+} = 1, W^{-} = 1)\).

\begin{proposition}[Preservation--Correction Frontier]
\label{prop:frontier}
With accuracy of private impression \(q\in(\tfrac12,1)\), the welfare point depends on the AI
accuracy \(\alpha\) only through which side of the Gateway it falls on:
\[
(W^{+},W^{-})=
\begin{cases}
\bigl(1,\ 0\bigr), & \alpha\geq q,\\[6pt]
\left(\dfrac{q}{1-q+q^{2}},\ \dfrac{q^{2}}{1-q+q^{2}}\right), & \tfrac12<\alpha<q,
\end{cases}
\]
where the two cases correspond to the AI falling above and below the Gateway,
respectively.
\end{proposition}

Within each regime \((W^{+},W^{-})\) is constant in \(\alpha\), and the two
regime points are mutually non-dominating: the above-Gateway point has strictly
higher preservation \(\bigl(1>\tfrac{q}{1-q+q^{2}}\bigr)\) and strictly lower
correction \(\bigl(0<\tfrac{q^{2}}{1-q+q^{2}}\bigr)\). Raising AI accuracy thus trades preservation against correction rather than improving both. The question of when to let the AI lead then becomes:



\begin{remark}[When should the AI lead?]
\label{rem:bangbang}
Under an asymmetric loss that weights a wrong cascade at $\kappa$ times the cost of a missed truth, welfare collapses to a weighted accuracy and letting the AI lead is optimal if and only if
\[
\pi>\pi^{*}(\kappa)=\frac{\kappa q^{2}}{(1-q)^{2}+\kappa q^{2}}.
\]
At $\kappa=1$ (equal costs) this reduces to the unconditional-accuracy threshold $\pi^{*}(1)=\frac{q^{2}}{1-2q+2q^{2}}$; as $\kappa$ grows, $\pi^{*}(\kappa)$ rises toward $1$---the costlier an amplified falsehood, the more accurate the AI must be before it leads.
\end{remark}

\begin{figure}[htbp]
  \centering
  \includegraphics[width=\linewidth]{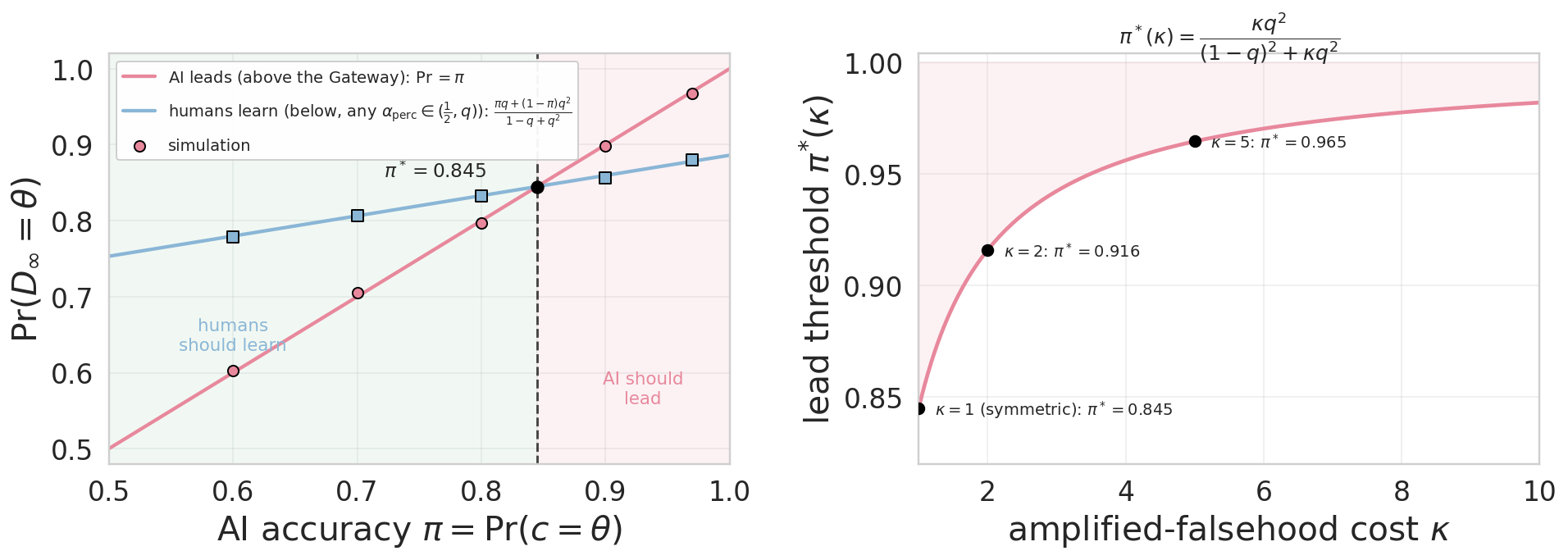}
  \caption{When should the AI lead? (Remark~\ref{rem:bangbang}, \(q = 0.7\).)
  \textbf{Left}: unconditional accuracy \(\Pr(D_\infty =
  \theta)\) against the AI's true accuracy \(\pi = \Pr(c = \theta)\), under the
  two regimes a deployment can realize at the same \(\pi\). If the population
  defers from the first user (above the Gateway), accuracy equals \(\pi\). If it
  keeps learning (below the Gateway, i.e.\ the indicator is weighted as one more
  piece of weak evidence), accuracy is \(\tfrac{\pi q + (1-\pi)q^{2}}{1-q+q^{2}}\)---%
  identical for every perceived accuracy \(\alpha_{\mathrm{perc}} \in
  (\tfrac{1}{2}, q)\) by Theorem~\ref{thm:direction}(b), and capped at
  \(\tfrac{q}{1-q+q^{2}} \approx 0.886\) no matter how accurate the AI truly is.
  The two cross at \(\pi^{*} \approx 0.845\); markers are simulation. \textbf{Right}: under an
  asymmetric loss where an amplified falsehood costs \(\kappa\) times a missed
  truth, the threshold \(\pi^{*}(\kappa) = \kappa q^{2}/((1-q)^{2} + \kappa
  q^{2})\) rises from \(0.845\) (\(\kappa = 1\)) toward \(1\).}
  \label{fig:welfare-threshold}
\end{figure}

Figure~\ref{fig:welfare-threshold} plots both cases at $q=0.7$. Figure~(a) compares unconditional accuracy across the two regimes: above the Gateway it equals $\pi$; below it the curve $\tfrac{\pi q+(1-\pi)q^{2}}{1-q+q^{2}}$ rises more slowly, crossing at $\pi^{*}(1)\approx 0.845$---a $90\%$-accurate AI qualifies---and the same true accuracy can realize either regime through the perceived strength of the indicator (Section~\ref{sec:emp-relax}). Figure~(b) shows $\pi^{*}(\kappa)$ rising from $0.845$ toward $1$~\footnote{Simulations that confirm every closed form of this section to simulation error; the per-theorem sweeps and figures are in the supplementary material.}.

\subsection{Escaping the frontier: structural levers from the theory}
\label{sec:interventions}

The frontier of Proposition~\ref{prop:frontier} constrains a single shared indicator: raising AI accuracy trades preservation for correction, never lifting both. Escaping it requires changing what is deployed. We develop two structural levers here and simulate both in Section~\ref{sec:sim-interventions}; further levers---including what to report alongside the consensus and when to attach the indicator---are examined in the supplementary material.

\noindent\textbf{Diversify the model (what to deploy).}
Replacing the shared \(c\) with independent per-user draws \(c_1,\dots,c_N\) breaks the common-signal dependence, so the information ceiling rises beyond the single term \(I(c;\theta)\):
\[
I(\Jbar_N;\theta)
\;\leq\;
I\bigl((c_1,\ldots,c_N,\sigma_{<\tau});\theta\bigr),
\qquad
I\bigl((c_1,\ldots,c_N);\theta\bigr)\uparrow\ \text{in } N.
\]
A cascade still forms eventually, so aggregation saturates at a higher but still \(N\)-independent level; full aggregation is approached only when the diversified signal is weighted strongly enough to keep the chain out of a cascade. Section~\ref{sec:sim-interventions} quantifies the gain.

\noindent\textbf{Trigger expert review at slow cascades (beyond deploying AI).}
By Proposition~\ref{prop:tau}(c), an incorrect AI prediction takes longer to cascade than a correct one,
\[
\E[\tau\mid c\neq\theta]\;>\;\E[\tau\mid c=\theta],
\]
so a large realized \(\tau\) is an observable symptom of \(c\neq\theta\)---a symptom that exists only because the AI is present. Triggering expert review when \(\tau\) exceeds a threshold \(\tau^*(q)\) therefore directs a scarce perfect signal selectively toward wrong cascades without observing \(\theta\), whereas an unconditional expert override corrects direction but, like any strong signal, restores no aggregation.

%% file: sections/06-empirical-calibration.tex
\section{Empirical Calibration of the Theoretical Analysis}
\label{sec:empirical}

The analysis so far characterizes calibrated Bayesian users, where users know the true accuracy of both their own impressions and the AI indicator, and update
beliefs rationally. We now calibrate the theoretical analysis by confronting the assumption using  real human-subject experiment data. The calibration is organized around three questions:
(1) \textit{Benchmark} Do users behave as calibrated Bayesians?
(2) \textit{Pooled calibration}: How does the average user weight different signals?
(3) \textit{Heterogeneity} : 
  Are there distinct types of users who weigh signals differently? 
We use this section to answer the three questions respectively, and leave the residual behavioral patterns not captured by these three
questions in the supplementary material. 


\subsection{Empirical Calibrations}
\label{sec:setting}
\noindent\textbf{Empirical data from real human-subject experiments}
We use a public dataset from prior work conducting a human-subject experiment on
misinformation spread in the presence of AI-based credibility indicators~\cite{lu2022effects}. The
dataset contains \(480\) sequences of news veracity judgments. In each sequence,
\(11\) participants review a news item, make an initial binary veracity judgment
(i.e., fake or real), review the public history, and then make their final
veracity judgment \(J_i\). Each final judgment is then appended to the public
history. We use this dataset because it closely follows the setting of information
cascade in the presence of an AI-based credibility indicator as a public signal.
Crucially, participants' initial judgments before reviewing the public history
are separately recorded, giving a direct read of the private impression \(s_i\).
For more details, please refer to the supplementary material.

The experiment has three between-subjects treatments of \(160\) rooms
each: \textsc{Control} (no AI; the classical BHW limit
\(\alpha=\tfrac{1}{2}\)), \textsc{AI-before} (AI shown before the participants'
impression and the public history), and \textsc{AI-after} (AI shown after). The treatments
share the same \(40\) news items and hold \(\alpha\) fixed by design, with each participant reviewing the same news story at most once. The empirical AI accuracy is \(\widehat\alpha=0.750\), and the
private-impression accuracy, identified from its pilot study, is
\(\widehat q\approx0.515\).

\subsection{Calibration in Three Readings}
\label{sec:calibration}\label{sec:emp-cal}\label{sec:emp-behavioral}

\subsubsection{Q1 (Benchmark): fully calibrated Bayesian users}
\label{sec:strict}\label{sec:emp-strict}

This benchmark is intentionally strict. It is a reference point, not a behavioral model of participants: it shows how observed reliance departs from calibrated Bayesian updating. Suppose users are fully calibrated Bayesian agents who know both their own
accuracy and the AI model's accuracy. Then the empirical AI accuracy
\(\widehat\alpha\) and the private-impression accuracy \(\widehat q\) can be used directly determine the dynamics. Thus, users should lie above the Gateway: the AI prediction should be strong enough that no private impression is
revealed, and all users' final judgments should match the AI prediction on each
news story. However, the empirical data depart from this strict benchmark in several ways: users do
not always follow the AI, full-room agreement with \(c\) is rare, consensus does not perfectly track AI correctness, and the realized information plateau remains
below the strong-AI ceiling \(I(c;\theta)=0.131\). Full diagnostics are reported in the supplementary material. These gaps show that the fully calibrated
assumption is too strong as a behavioral description.

Meanwhile, the main qualitative pattern is still visible. The AI
prediction significantly affects the consensus direction: when the AI is wrong, both AI treatments fall well below the \(50\%\) truth-accuracy line, while \textsc{Control} stays roughly symmetric around \(0.54\). Collective information also saturates, with
limited wisdom-of-crowds growth in \(n\). The following subsections relax the decision rule to close this gap.

\subsubsection{Q2 (Calibration): behavioral weights of an average user}
\label{sec:pooled}

A growing line of work in human-AI interaction suggests relaxing the decision rule, where users may weight the signals by their own perceived reliabilities instead of calibrated probabilities. We estimate, directly from the choices, the per-signal log-odds each signals carries for the user,
\[
\ell_i^\beta
=
\ell_0
+
\beta_{\mathrm{private}}\,s_i
+
\beta_{\mathrm{AI}}\,c_i
+
\beta_{\mathrm{public}}\,T_i,
\]
with \(s_i, c_i \in \{-1,+1\}\) and \(T_i=\sum_{j<i}(2J_j-1)\) the signed public tally, so that \(\beta_{\mathrm{AI}}\) is the perceived log-odds the user assigns to one AI indication, \(\beta_{\mathrm{private}}\) to her own impression, and \(\beta_{\mathrm{public}}\) to one net peer judgment. Because the logit temperature is not separately identified, we read only ratios of signal weights: the AI-private ratio is identified from \textsc{AI-after}, where the initial judgment is recorded before the AI and the public history, while \textsc{AI-before} serves only to compare the AI signal against the public-history weight, since its recorded initial judgment may already carry the AI's influence. Two points are essential. First, these weights are identified from the choice data alone: unlike the Q1 benchmark, the estimation does not use the empirical accuracies \(\widehat\alpha,\widehat q\), and every conclusion below is a ratio of these coefficients---hence invariant to \(\widehat\alpha,\widehat q\). Second, the behavioral first-user threshold compares the perceived AI log-odds against the perceived self log-odds, \(\beta_{\mathrm{AI}}\) versus \(\beta_{\mathrm{private}}\):

\begin{proposition}[Behavioral threshold for first-user cascade]
\label{prop:behavioral}
In the behavioral model, the first-user cascade analog of
Theorem~\ref{thm:cascade-genericity} holds with \(\alpha\geq q\) replaced by the
requirement that the user assign the AI indication at least as much log-odds as
her own impression,
\[
\rho_{\text{AI-private}} \;\equiv\; \frac{\beta_{\mathrm{AI}}}{\beta_{\mathrm{private}}} \;\geq\; 1
\qquad\bigl(\text{equivalently } \beta_{\mathrm{AI}} \geq \beta_{\mathrm{private}}\bigr).
\]
When this holds, \(J_i=c\) for all \(i\geq1\); otherwise the chain starts in the
learning region, with starting point scaled by \(\rho_{\text{AI-private}}\). The
threshold compares perceived weights only and does not involve the
signals' true accuracies; the equivalent effective perceived accuracy at which
the user reaches the Gateway is \(\alpha^{*}(q,\rho_{\text{AI-private}})\),
solving \(L_{\alpha^{*}}=\rho_{\text{AI-private}}^{-1}\,\Lq\).
\end{proposition}

Since the behavioral weights are estimated under a logistic choice rule, an
estimated \(\rho_{\text{AI-private}}\geq 1\) places a user on the cascading side
of the behavioral Gateway; the immediate cascade itself is the deterministic
limit of that rule.




We first estimate a pooled decision rule. Users' judgments in \textsc{Control} are used only to estimate how users weigh the public history relative to their own initial judgment when no AI signal is present. Then, the two experimental treatments provide two settings for the estimation. Recall that in the \textsc{AI-after} treatment, each user's initial judgment is recorded before she sees any other signal, so it can be used directly as \(s_i\); this lets us estimate the
AI-private ratio $\rho_{\text{AI-private}}$.  \textsc{AI-before}, however, serves a different purpose. There, the subject sees the AI before forming the recorded initial judgment, so that initial judgment may already be
influenced by the AI and cannot be treated as a clean \(s_i\). We instead use it only to compare the
AI-versus-public ratio $\rho_{\text{AI-public}} \equiv \frac{\beta_{\mathrm{AI}}}{\beta_{\mathrm{public}}}$. All confidence intervals reported below are subject-level bootstrap intervals, resampling subjects with replacement, each entering with all of her judgments across rooms. Full specification and robustness checks are reported in the supplementary material.

The estimates show that under the behavioral relaxation, the average user does not lie above the Gateway. In \textsc{AI-after}, \(\rho_{\text{AI-private}} = 0.49\) (subject-bootstrap \(95\%\) CI
\([0.33,0.67] < 1\)), so the
average user gives one AI prediction about half the weight of one private
impression. By Proposition~\ref{prop:behavioral}, this places the average user in
the interior learning regime, which explains the gap in Q1: users should not all follow the AI from the first position onward, but should instead show partial AI alignment and partial tracking of AI correctness, as in the data. The AI-based credibility indicator is nonetheless influential: in \textsc{AI-after} one AI prediction carries the weight of roughly three prior
peer judgments (\(\rho_{\text{AI-public}} \approx 3.5\)), even though it weighs less than the user's own private
impression. Here a ``peer judgment'' means one unit of the signed public tally $T_i$ in the behavioral choice model, not an independently observed private impression. This pull is stronger when the AI is shown first: under the same reduced-form specification, \(\rho_{\text{AI-public}}\) is \(6.1\) in \textsc{AI-before} versus \(3.5\) in \textsc{AI-after}, so presenting the AI before the user forms her own
impression anchors judgment harder than showing it afterward. Against the strict theory this is a steep discount. Because measured private-impression accuracy is close to chance, \(\widehat\Lq\) is small, so the fully calibrated Bayesian benchmark assigns the AI a very large relative weight---at the point estimate \(k^{*}=\widehat\La/\widehat\Lq=1.099/0.061\approx 18\) peer judgments (Remark~\ref{rmk:equivalent-population}). The behavioral estimate is far smaller, \(\rho_{\text{AI-public}}\approx 3\)--\(6\) peer judgments---the discount that keeps the average user learning.

\begin{figure}[htbp]
  \centering
  \includegraphics[width=\linewidth]{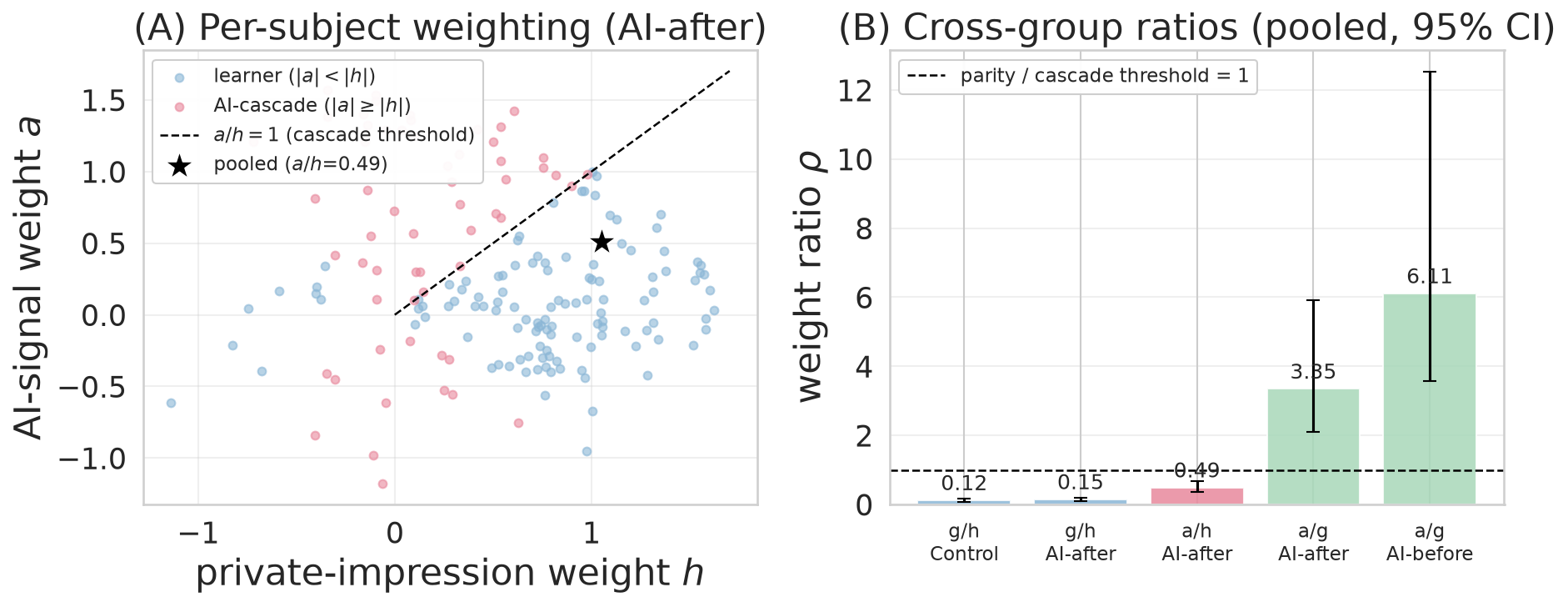}
  \caption{Subject-level behavioral heterogeneity from the EM mixture of logistic regressions. \textbf{(A)} Each point is one \textsc{AI-after} subject's AI-signal weight \(\beta_{\mathrm{AI}}\) against her private-impression weight \(\beta_{\mathrm{private}}\). Points above the threshold \(\beta_{\mathrm{AI}}=\beta_{\mathrm{private}}\) place more weight on the AI than on their own impression and therefore lie on the behavioral cascading side of the Gateway. Points below the threshold remain in the learning region. \textbf{(B)} The weight ratios \(\rho\) with subject-bootstrap \(95\%\) CIs. The pooled ratio \(\rho_{\text{AI-private}}=0.49\) lies below parity, but the subject-level mixture reveals a cascade-prone minority with \(\rho_{\text{AI-private}}>1\).}
  \label{fig:behavioral-em}
\end{figure}

\subsubsection{Q3 (Heterogeneity): behavioral weights across users}
\label{sec:mixture}

The pooled estimate, however, masks within-population heterogeneity: the data contain sequences that look like first-user AI-led cascades, with the first user abandoning her private impression and later judgments highly aligned with the AI prediction. The distributional question is whether the AI creates a cascade-prone behavioral component that cascades on the indicator alone even when the average user does not. We address this by fitting a \(K\)-type mixture of logistic decision rules over subjects via EM, with \(K\) selected by BIC (specification in the supplementary material).

The selected \(K=3\) mixture reveals substantial heterogeneity. About \(46\%\) of subjects weigh their own impression well above the AI, about \(20\%\) are assigned to a component whose fitted weights place at least as much weight on the AI as on their own impression (\(\rho_{\text{AI-private}}\geq 1\)), the cascading side of the behavioral Gateway, and the remaining \(\approx 34\%\) are not cleanly classified into either regime. The comparison with \textsc{Control} makes this point sharper. In \textsc{Control}, the mixture does not identify an individual crowd-cascading class. This is consistent with the classical BHW logic: a single peer judgment is not enough to overturn one's
own private impression, and cascades form only through accumulated public
history. The AI is different because it enters as a strong common signal. Even
when the average user discounts it, some users treat it as strong enough to
override their own impression.

Taken together, the three readings provide behavioral evidence consistent with the shared-signal mechanism: the average user weighs the AI below her own impression, while a cascade-prone behavioral component crosses the behavioral Gateway and cascades on the indicator alone. These conclusions are stated as ratios because the logit temperature is not identified; \(\rho_{\text{AI-private}}\) is identified only in \textsc{AI-after}, while \textsc{AI-before} is used only for the timing comparison through \(\rho_{\text{AI-public}}\).

%% file: sections/07-simulation.tex
\section{Simulations Extended from the Calibration}
\label{sec:simulation}

The empirical calibration shows a gap between rational Bayesian users and user behavior in the wild. We close the paper with simulation in two parts. First we chart miscalibrated trust, the lever a platform's presentation controls and the one the calibration already measured (Q2). Second, we briefly illustrate two structural levers the theory suggests (Section~\ref{sec:interventions})---diversifying the model and cascade-triggered expert review---across the two regimes ($\alpha > q$ and $\alpha < q$) and across populations that under- or over-weight the AI. Simulation results of remaining deployment levers are deferred to the supplementary material.

\subsection{Miscalibrated trust: the perceived Gateway}
\label{sec:perceived}\label{sec:emp-relax}

\begin{figure}[htbp]
  \centering
  \includegraphics[width=\linewidth]{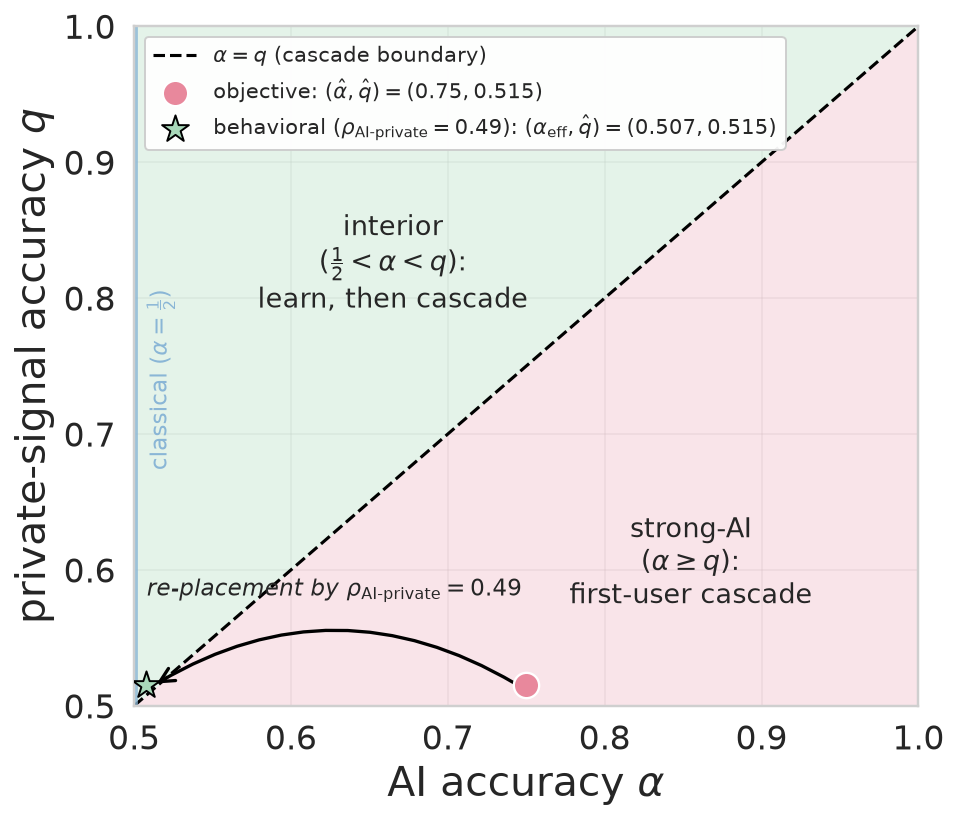}
  \caption{Regime structure on the $(\alpha, q)$ plane (symmetric prior). The $\alpha = q$ diagonal separates the interior regime (learn, then cascade) from the strong-AI regime (first-user cascade); $\alpha = \tfrac{1}{2}$ is the classical BHW edge. The same experimental population appears under two placements (Section~\ref{sec:calibration}): its objective calibration $(\widehat\alpha, \widehat q) = (0.75, 0.515)$ sits deep in the strong-AI regime, while its behavioral placement---the effective AI strength backed out from choices---falls across the boundary into the interior. The gap between the two points is the main empirical finding of Section~\ref{sec:empirical}.}
  \label{fig:phase-diagram}
\end{figure}

In the model the Gateway is written in true signal accuracies; in deployment the operative Gateway is behavioral, because users may weight the AI as if it had a different reliability. The Gateway comparison runs on the accuracy users believe, not the
accuracy that generates $c$---and the behavioral weights Section~\ref{sec:empirical}
estimated (Q2) are exactly a measurement of this gap; we chart the consequences
here. Figure~\ref{fig:phase-diagram} fixes the canvas: the $(\alpha, q)$ plane
with the $\alpha = q$ diagonal separating the interior regime from the strong-AI
regime; Section~\ref{sec:empirical} placed the experimental population at two
points on it---once at its measured accuracies, once at its behavioral weights.

\begin{figure}[htbp]
  \centering
  \includegraphics[width=\linewidth]{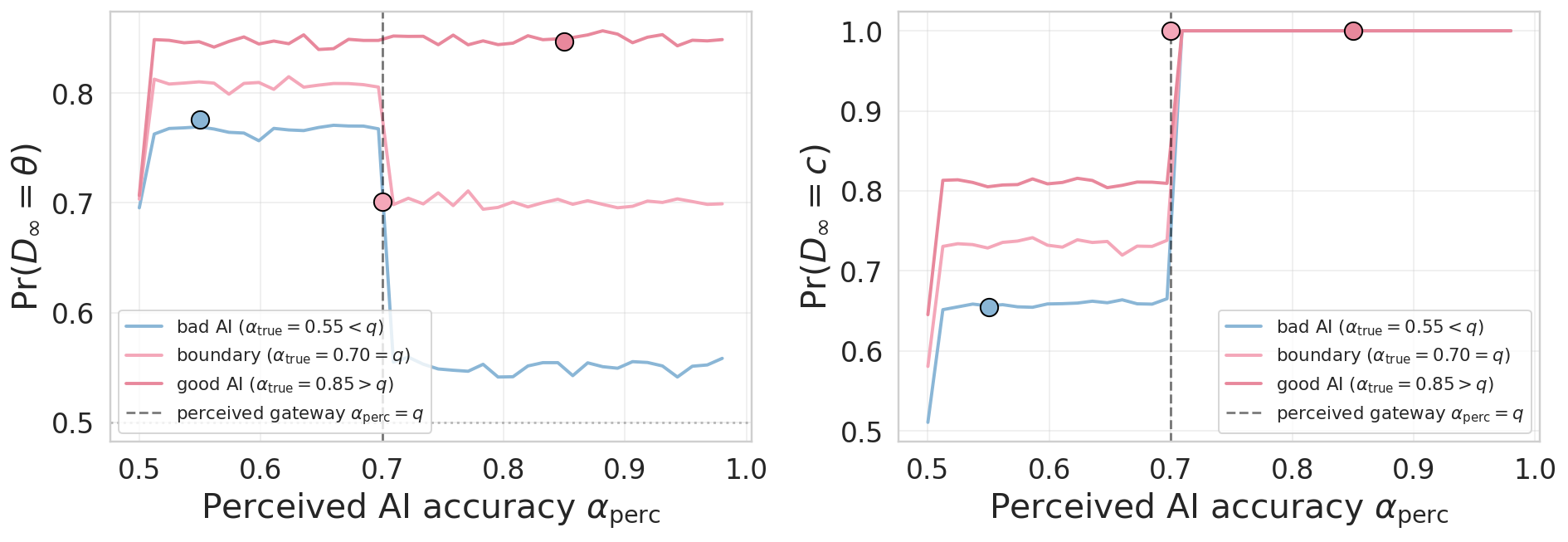}
  \caption{Miscalibrated trust: users update on a perceived accuracy $\alpha_{\mathrm{perc}}$ while $c$ is generated with true accuracy $\alpha_{\mathrm{true}}$ ($q = 0.7$, $N = 50$, $12{,}000$ simulated cascades per point). \textbf{Right}: $\Pr(D_\infty = c)$ jumps to $1$ at the perceived Gateway $\alpha_{\mathrm{perc}} = q$, identically for every $\alpha_{\mathrm{true}}$. \textbf{Left}: $\Pr(D_\infty = \theta)$ is set by $\alpha_{\mathrm{true}}$; over-trusting a weak AI ($\alpha_{\mathrm{true}} = 0.55$) collapses it from $\approx 0.77$ to $\approx 0.55$. Dots mark the calibrated case $\alpha_{\mathrm{perc}} = \alpha_{\mathrm{true}}$.}
  \label{fig:relax-calibration}
\end{figure}

\begin{figure}[htbp]
  \centering
  \includegraphics[width=\linewidth]{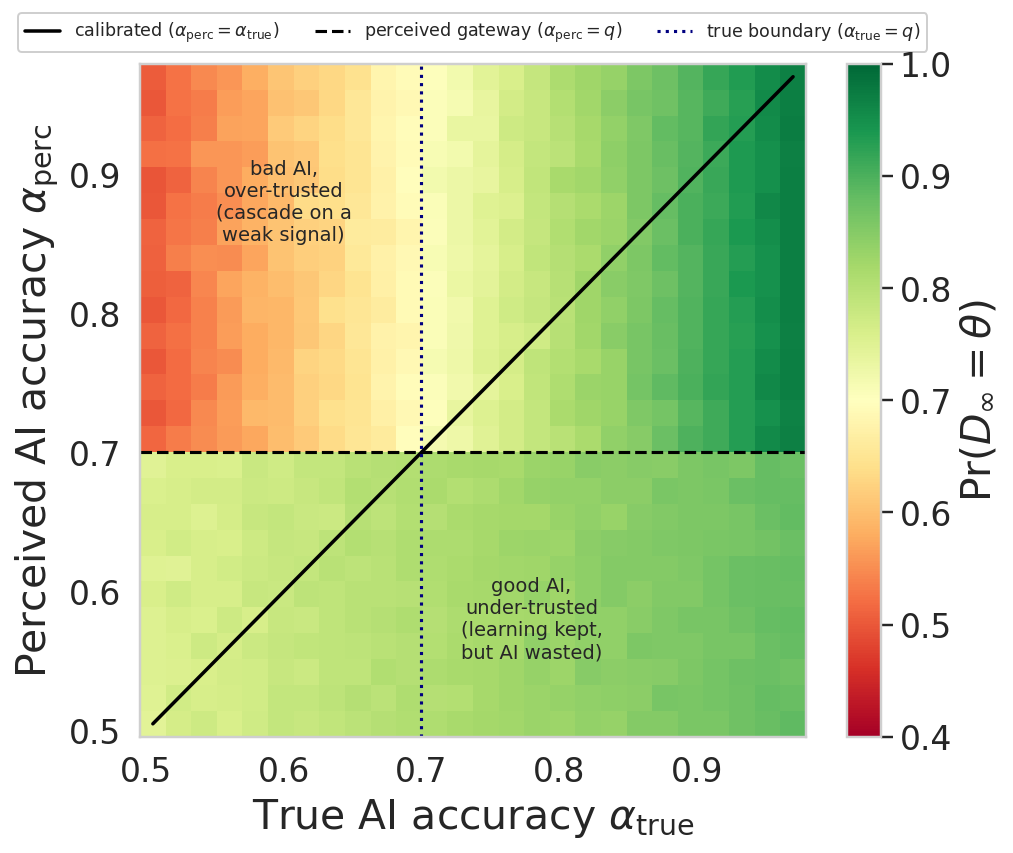}
  \caption{$\Pr(D_\infty = \theta)$ over the $(\alpha_{\mathrm{true}}, \alpha_{\mathrm{perc}})$ plane ($q = 0.7$, $N = 40$). Above the perceived Gateway $\alpha_{\mathrm{perc}} = q$ (dashed) the cascade direction follows $c$ and accuracy reduces to $\alpha_{\mathrm{true}}$ (vertical bands); below it private impressions continue to enter the public history. The diagonal is the calibrated case; the top-left corner is a weak AI over-trusted, the bottom-right a strong AI under-trusted.}
  \label{fig:relax-region}
\end{figure}

To separate belief from truth, we let users update on a perceived accuracy
$\alpha_{\mathrm{perc}}$ while $c$ is drawn with true accuracy
$\alpha_{\mathrm{true}}$ (symmetric prior, $q = 0.7$). The two parameters
control different things (Figure~\ref{fig:relax-calibration}): the cascade
direction follows the AI prediction---$\Pr(D_\infty = c) = 1$---exactly
when $\alpha_{\mathrm{perc}}$ crosses $q$, identically for every
$\alpha_{\mathrm{true}}$, while the accuracy of that direction,
$\Pr(D_\infty = \theta)$, is set by $\alpha_{\mathrm{true}}$. Sweeping the
whole $(\alpha_{\mathrm{true}}, \alpha_{\mathrm{perc}})$ plane
(Figure~\ref{fig:relax-region}) indicates the resulting asymmetry: over-trusting
a weak AI carries the population across this perceived Gateway onto a
near-chance signal, pushing $\Pr(D_\infty = \theta)$ below even the no-AI
level, while under-trusting a strong AI only leaves its information unused.
The asymmetry this section illustrates---over-trusting a weak indicator is
worse than having none, under-trusting a strong one merely wastes it---is
the simulated counterpart of what the data now measure.

\subsection{Exploratory Analysis: simulations of AI diversification and expert review interventions}
\label{sec:sim-interventions}

We conduct an exploratory analysis by simulating two structural interventions described in
Section~\ref{sec:interventions}: diversification and expert review. We compare
both against the shared AI prediction baseline across strong and weak AI regimes
\((\alpha=0.85>q\) and \(\alpha=0.60<q\), with \(q=0.7\) and \(N=60)\), while
varying the behavioral weight placed on the AI. Both interventions can improve
welfare relative to a shared AI prediction, but through different mechanisms.
Diversification replaces the shared signal with independent AI predictions across
users, breaking the common-dependence problem. Expert review instead uses slow
cascade formation as a warning signal: when the crowd remains below the Gateway,
a slow cascade is informative that the AI may be wrong, so expert review can
redirect the consensus.

The welfare comparison suggests the main tradeoff. In these simulations,
diversification is the more robust intervention: it remains no worse than the no-AI condition, prevents the
collapse that occurs when a weak shared indicator is over-trusted, and moves
outcomes toward the ideal \((1,1)\). In the over-trusting case, it raises
correction \(W^{-}\) from \(0\) to about \(0.97\) without sacrificing
preservation. Expert review can also be highly effective below the Gateway,
reaching roughly \((1,0.7)\), but its benefit depends on cascades forming slowly
enough to trigger review. When users over-weight the AI, cascades form
immediately, so \(\tau\) no longer carries useful warning information and the
expert is rarely called. Expert review is therefore weakest exactly where the
shared-indicator baseline is most dangerous.

The heterogeneous simulation gives the empirically relevant case. Using a mixed
crowd motivated by Q3---\(60\%\) under-weighting users with
\(\beta_{\mathrm{AI}}=0.4\) and \(40\%\) over-weighting users with
\(\beta_{\mathrm{AI}}=2.5\)---diversification remains strong: for the strong AI,
it raises accuracy from \(0.85\) to \(0.99\) and \(W^{-}\) from \(0.19\) to
\(0.99\); for the weak AI, it raises accuracy from \(0.77\) to \(0.88\) and
\(W^{-}\) from \(0.55\) to \(0.83\). Expert review degrades because the
over-weighting minority often induces immediate cascades, making slow formation a
weaker trigger; for the strong AI, the trigger fires in only \(13\%\) of rooms and
raises \(W^{-}\) only to \(0.31\). Thus both interventions can help, but within our simulated settings
diversification is more reliable across AI strengths and behavioral weightings,
while expert review is useful mainly for still-learning crowds.

\begin{figure}[htbp]
  \centering
  \includegraphics[width=\linewidth]{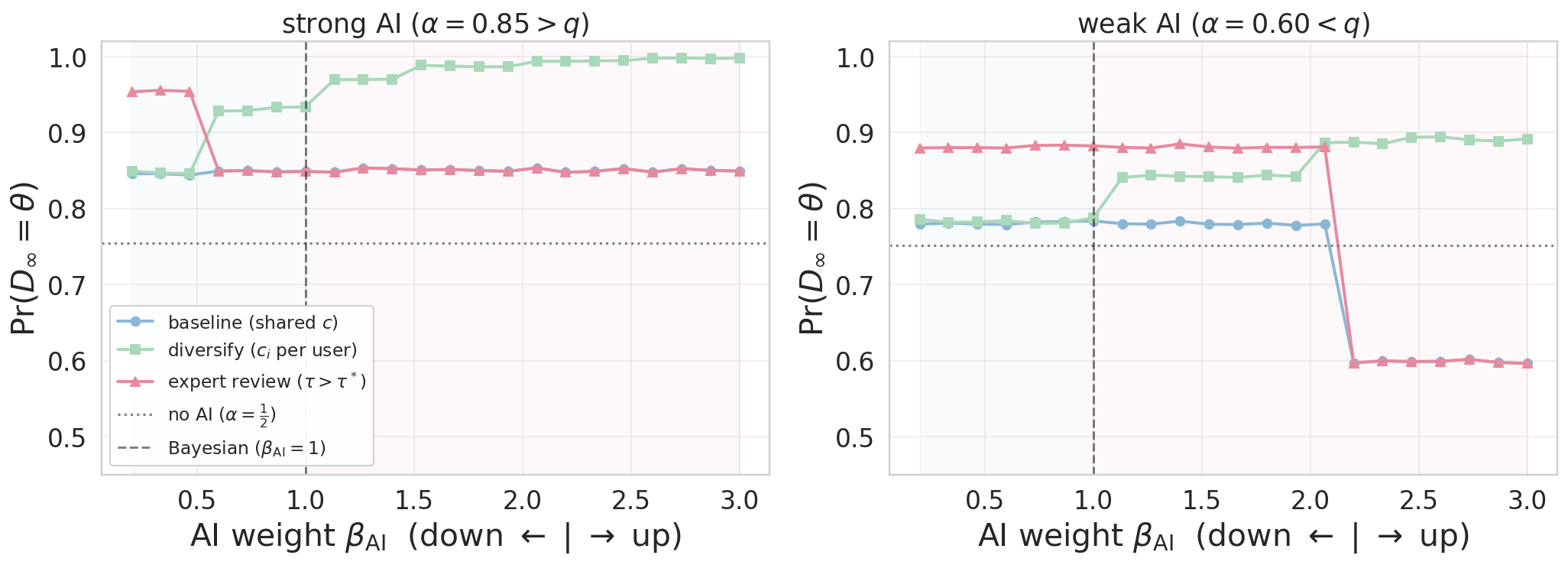}
  \caption{Consensus accuracy $\Pr(D_\infty = \theta)$ against the behavioral AI weight $\beta_{\mathrm{AI}}$ (down-weighting $\leftarrow$ | $\rightarrow$ up-weighting), for a strong AI ($\alpha = 0.85 > q$, left) and a weak AI ($\alpha = 0.60 < q$, right); $q = 0.7$, $N = 60$, $30{,}000$ cascades per point. \textbf{Baseline} (shared $c$) is flat at $\alpha$ above the Gateway and crashes below the no-AI line when a weak indicator is over-trusted. \textbf{Diversify} ($c_i$ per user) never falls below no-AI and rises toward $1$ as the AI is weighted more. \textbf{Expert review} ($\tau > \tau^{*}$, best $\tau^{*}$) lifts accuracy only below the behavioral Gateway; once $\beta_{\mathrm{AI}}$ crosses it the first user cascades ($\tau = 1$), the trigger never fires, and the lever collapses to the baseline.}
  \label{fig:int-outcomes}
\end{figure}

\begin{figure}[htbp]
  \centering
  \includegraphics[width=\linewidth]{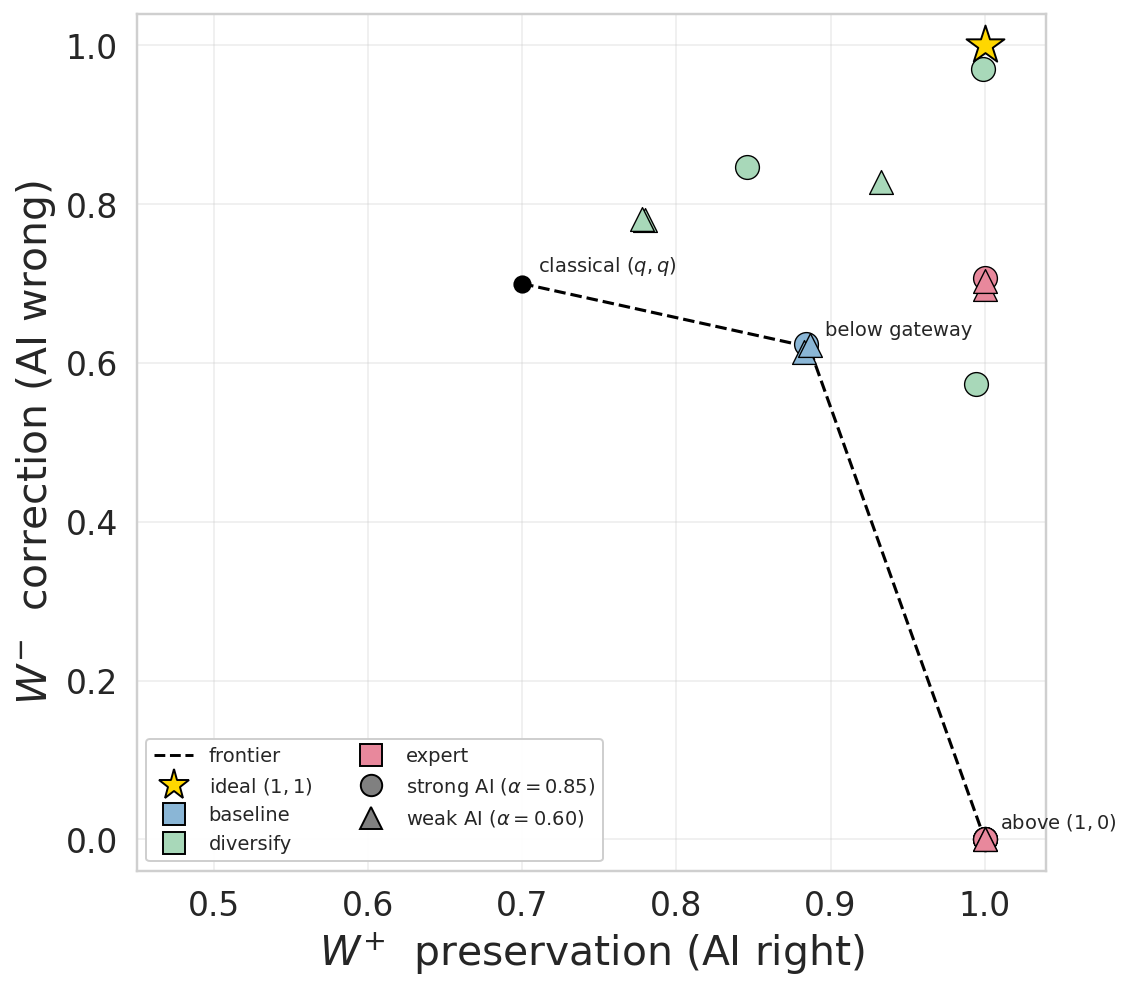}
  \caption{Welfare plane $(W^{+}, W^{-})$ of Proposition~\ref{prop:frontier} for each regime ($\alpha = 0.85$ circles, $\alpha = 0.60$ triangles) and weighting (down $\beta_{\mathrm{AI}} = 0.4$, Bayesian $1.0$, up $2.5$); $q = 0.7$, $N = 60$. The baseline (blue) sits on the AI-strength frontier. Diversify (green) moves toward the ideal $(1,1)$ across the grid---up to $(1.00, 0.97)$ for an over-trusting strong-AI population. Expert review (pink) reaches $(1, \approx\!0.7)$ below the Gateway but is pinned at the above-Gateway $(1,0)$ under over-reliance, where its slow-cascade trigger never fires.}
  \label{fig:int-welfare}
\end{figure}

%% file: sections/08-discussion.tex
\section{Discussion}\label{sec:discussion}
In this study, we introduce a lens for characterizing how people
interact with AI-based credibility indicators during misinformation spread. We
extend the classic Bayesian belief-updating model by treating the AI indicator
as a shared public signal. Through theoretical analysis, empirical calibration,
and simulations, we show how users compare their private impressions with AI
predictions, and how this comparison shapes collective reliance on AI and welfare
in misinformation judgments. We now discuss the implications, limitations, and
future directions of this work.

\subsection{Theoretical implication}

Our use of social learning theory is deliberately conservative: we take the classical framework and point it at a contemporary platform intervention. AI credibility indicators differ from ordinary advice because they are deployed as shared signals---the same output reaches many users, and its influence can be mistaken for independent crowd agreement. That distinction is easy to miss in individual-level human-AI studies, and becomes central once AI assistance is embedded in sequential social systems.

Our theoretical contribution is a lens: casting the AI indicator as a \emph{common public signal} traces its consequences through the same belief-update dynamics that govern the crowd. Unlike signals that arise from the population (prices, ratings, announcements), the AI prediction is exogenous and drawn once per story; it enters every posterior as the same constant term, relocating where the public process starts (Theorem~\ref{thm:equilibrium}) rather than adding a step to it. The failure this creates---bounded crowd information, collapsing to the AI signal alone above the Gateway---is distinct from the bounded-belief~\cite{smith2000pathological} and sparse-network~\cite{acemoglu2011bayesian} obstructions, and is the discrete analog of correlated signals overstating an unweighted average's precision.

Two features sharpen the mechanism. First, the Gateway is a property of the human--AI pair, not the AI alone: it runs on the accuracy users believe (Proposition~\ref{prop:behavioral}), so one indicator can sit in different regimes for different populations---exactly what the calibration finds, the experimental population being objectively strong-AI yet behaviorally interior, the gap between the two its central empirical result. Second, no single accuracy number evaluates the system: by Proposition~\ref{prop:frontier} the welfare point depends on $\alpha$ only through which side of the Gateway it falls on, and the two regime points are mutually non-dominating---a higher $\alpha$ buys preservation $W^{+}$ at the cost of correction $W^{-}$---so a credibility indicator in a sequential environment must be assessed on both coordinates.

\subsection{Implications for human--AI interaction in information spread}

Our results suggest that AI-based credibility indicators change the role of
public history in information spread. The public history is not irrelevant: users
still observe earlier judgments and update on them. What changes is what those
judgments mean. Without a shared AI signal, early judgments can reveal private
impressions and help the crowd aggregate information. With a shared AI signal,
the same public history may instead record repeated responses to a common input.
A large AI-aligned majority can therefore look like independent human validation
while containing little independent human evidence. This distinction matters for the design and interpretation of credibility
indicators. Platforms should not treat an AI-aligned majority as straightforward
evidence that many independent users validated the claim. The empirical
calibration shows that users do use the AI signal, sometimes strongly enough to
cascade on it, and that showing the AI before users form their own impressions
can strengthen this influence. Thus the public record produced after an AI
warning should be interpreted as a mixture of human impressions, social
influence, and shared AI dependence, not as a simple crowd statistic.

Therefore, platforms need to preserve or disclose the source
structure behind the public history, where similar patterns in recent research on community notes and user interaction visualization~\cite{feng2023examining,hughes2024viblio} should be further highlights when AI is presented. Communicating uncertainty and provenance can
help observers distinguish independent crowd evidence from repeated responses to
one AI signal. Meanwhile, simply canceling or delaying the AI is not enough, because a strong indicator can still redirect later users while leaving the reported majority difficult to interpret. More structural interventions, such as diversifying AI signals or triggering expert review when cascades form slowly, are better aligned with the problem because they change the information structure rather than only changing when or how strongly the same shared signal is shown.

\subsection{Implications for designing AI interventions against misinformation}

Our results suggest that misinformation interventions should not rely only on
tuning a single shared AI indicator, if it is not a perfect oracle. Changing the strength, timing, or presentation of the same indicator can move the system along the
preservation-correction frontier, but it does not remove the frontier itself, as
stronger reliance better preserves correct AI predictions, while making wrong
predictions harder for the crowd to correct. The design problem is therefore not
simply how to make people utilize AI, but how to keep the AI and the human crowd jointly informative.

This points to interventions that change the information structure behind the
public history. Provenance disclosure helps observers distinguish independent
human evidence from repeated responses to the same AI signal, preventing
spurious confidence from growing with crowd size. Diversifying AI signals across
users breaks common dependence more directly and is the most reliable intervention
in our simulations, including for over-trusting crowds, although its benefit is
limited when users still cascade immediately. Cascade-triggered expert review
uses slow cascade formation as a warning signal for likely AI errors, but works
mainly below the behavioral Gateway, where cascades remain slow enough to be
diagnostic. Evaluation should therefore report preservation and correction
separately, as the results of over-reliance and under-reliance, respectively.

\subsection{Limitations and future work}

Our work has several limitations. As a cascade model, it retains the standard scaffolding of social-learning theory: binary veracity, conditionally independent private impressions, and Bayesian or behaviorally quasi-rational users. These assumptions help isolate the mechanism, but they simplify modern misinformation settings, where claims are often partial, ambiguous, multi-dimensional, or evolving rather than simply true or false. The empirical analysis relaxes the strict Bayesian assumption through behavioral weights, but conditional independence and the exogeneity of the AI indicator remain active constraints. Item-level clustering in the data including rooms that cascade against the indicator regardless of its correctness suggests that correlated impressions matter in practice.

The empirical calibration should be read as evidence for the mechanism, not as a definitive population-level test. It reuses one public experiment with one platform paradigm, one news corpus, and one subject pool. Subjects also recur across rooms, so sharp inference must account for dependence across sequences. In addition, private-impression accuracy is estimated near chance, behavioral weights are identified only up to a logit temperature, and the AI-private ratio requires a treatment where private impressions are recorded before AI exposure. Dedicated experiments that vary AI accuracy, private-impression accuracy, ordering, and provenance would provide a cleaner causal test of the Gateway.

Finally, the paper adopts one theoretical lens: sequential Bayesian belief updating along a chain. This lens fits our question about how a shared AI signal reshapes social learning, but it leaves out network structure, strategic user responses, optimal disclosure, and platform-level intervention design. It also models the AI indicator mainly through accuracy, leaving aside AI-specific properties such as calibration, explanation, confidence display, model diversity, provenance, and perceived authority. Future work can extend the model along these dimensions and test interventions such as diversified model outputs, provenance disclosures, and cascade-triggered expert review in field or large-scale platform experiments.

%% file: sections/09-conclusion.tex
\section{Conclusion}\label{sec:conclusion}
This paper extends the classical Bayesian cascade model with the AI-based
credibility indicator as a shared public signal. The resulting Gateway condition
shows that AI changes what public history means---crowd agreement may reflect
accumulated independent human evidence, or repeated dependence on the same AI
prediction---creating a preservation--correction trade-off. Calibration on
human-subject data shows that the average user weights the AI below her own
impression but above several peer judgments, and simulations suggest that
over-reliance on a weak AI is especially harmful while diversifying AI signals
across users can better keep the crowd informative. We discuss implications for
understanding human-AI interaction in information spread and for designing
AI-based interventions against misinformation.